\documentclass[journal]{IEEEtran}

\usepackage[utf8]{inputenc}
\usepackage{subfiles}
\usepackage{amsfonts}
\usepackage{amsmath,nccmath}
\usepackage{bm,upgreek}
\usepackage{graphicx,pifont}
\usepackage[noadjust]{cite}
\usepackage{color}
\usepackage{comment}
\usepackage{authblk}
\usepackage[ruled]{algorithm}
\usepackage{algpseudocode,algorithmicx}
\usepackage{subcaption}
\usepackage{ulem}

\usepackage{amssymb}
\usepackage{mathtools}
\usepackage{amsthm}
\usepackage{nicefrac}
\usepackage{siunitx} 
\usepackage{commath}
\usepackage{url}
\newcommand{\argmin}{\arg\!\min} 

\usepackage{tikz,pgfplots,pgfplotstable}   
\usepgfplotslibrary{groupplots,fillbetween,colorbrewer,statistics}
\usetikzlibrary{intersections,calc,arrows,matrix,spy,pgfplots.statistics, pgfplots.colorbrewer}
\usepackage{color}
\definecolor{darkred}{rgb}{0.7,0,0}
\definecolor{darkgreen}{rgb}{0,0.5,0}
\definecolor{darkblue}{rgb}{0,0,0.7}
\definecolor{SkyBlue}{rgb}{0.53, 0.81, 0.92}
\pgfplotsset{compat=1.5.1, cycle list/Set1-3}
\usepackage{tikz-3dplot} 
\newcommand{\ie}{\textit{i.e.}~}

\newtheorem{lemma}{Lemma}[section]
\newtheorem{theorem}[lemma]{Theorem}
\newtheorem{proposition}[lemma]{Proposition}

\newcommand{\commentsymbol}{$\triangleright$}
\makeatletter
\newcommand{\LineComment}[2][\algorithmicindent]{\Statex \hspace{#1}\commentsymbol{} #2}
\makeatother

\def\vdotfill#1{\vtop to0pt{\null \dimen0=#1\baselineskip\advance\dimen0 by-.4ex 
   \kern-1.6ex \cleaders\hbox{\lower.4ex\vbox to1ex{}.}\vskip\dimen0 \vss}}
   
\algdef{SE}[FOR]{NoDoFor}{EndFor}[1]{\algorithmicfor\ #1}{\algorithmicend\ \algorithmicfor}%
\algdef{SE}[IF]{NoThenIf}{EndIf}[1]{\algorithmicif\ #1}{\algorithmicend\ \algorithmicif}%
\algdef{SE}[WHILE]{NoDoWhile}{EndWhile}[1]{\algorithmicwhile\ #1}{\algorithmicend\ \algorithmicwhile}%

\usepgfplotslibrary{statistics}
    \pgfplotsset{
        compat=1.3,
        my boxplot style/.style={
            boxplot,
            draw=black,
            solid,
            fill=white,
            mark=*,
            every mark/.append style={
                fill=gray,
            },
        },
    }
\let\oldAA\AA
\renewcommand{\AA}{\text{\normalfont\oldAA}}

\title{Joint Angular Refinement and Reconstruction \\ for Single-Particle Cryo-EM}
\author[1,2]{Mona Zehni}
\author[1]{Laur\`ene Donati}
\author[1,3]{Emmanuel Soubies}
\author[2]{Zhizhen J. Zhao}
\author[1]{Michael Unser}
\affil[1]{Biomedical Imaging Group, \'Ecole polytechnique fédérale de Lausanne (EPFL), Switzerland}
\affil[2]{Coordinate Science Laboratory, University of Illinois at Urbana-Champaign, USA}
\affil[3]{IRIT, Universit\'e de Toulouse, CNRS, France}

\begin{document}

\maketitle

\begin{abstract}
Single-particle cryo-electron microscopy (cryo-EM) reconstructs the three-dimensional (3D) structure of bio-molecules from a large set of 2D projection images with random and unknown orientations. A crucial step in the single-particle cryo-EM pipeline is 3D refinement, which resolves a high-resolution 3D structure from an initial approximate volume by refining the estimation of the orientation of each projection. In this work, we propose a new approach that refines the projection angles on the continuum. We formulate the optimization problem over the density map and the orientations jointly. The density map is updated using the efficient alternating-direction method of multipliers,  while the orientations are updated through a semi-coordinate-wise gradient descent for which we provide an explicit derivation of the gradient. Our method eliminates the requirement for a fine discretization of the orientation space and does away with the classical but computationally expensive template-matching step. Numerical results demonstrate the feasibility and performance of our approach compared to several baselines.
\end{abstract}


\section{Introduction}

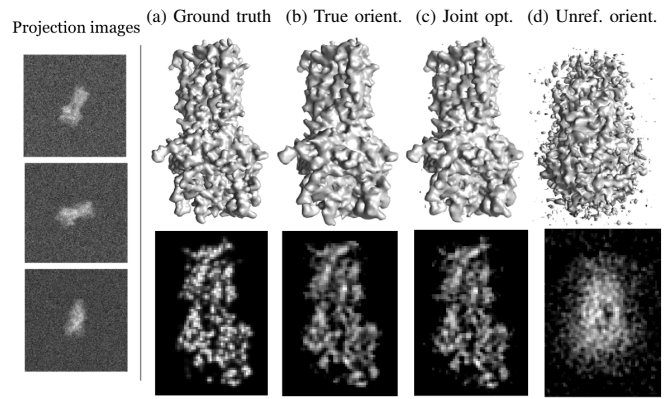
\begin{figure}[t!]
    \centering
    \begin{tikzpicture}
    \node at (0,0) {\hspace{-0.2cm}\includegraphics[width =  \linewidth]{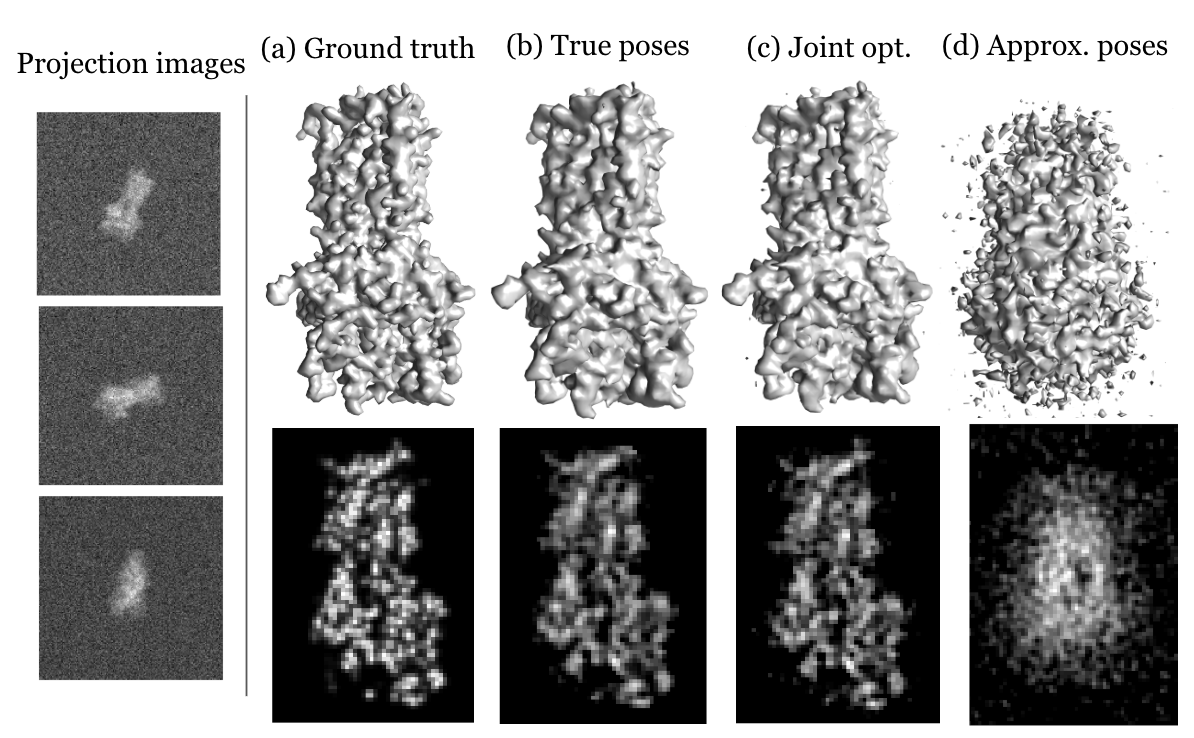}};
    \draw[white,fill=white] (-2.7,2.2) rectangle (4.5,2.6);
    \node at (-1.8,2.4) {\scriptsize (a) Ground truth};
    \node at (0,2.4) {\scriptsize (b) True orient.};
    \node at (1.6,2.4) {\scriptsize (c) Joint opt.};
    \node at (3.3,2.4) {\scriptsize (d) Unref. orient.};
    \end{tikzpicture}
    \caption{Single particle analysis. Left: Samples of the noisy projection images used in this experiment. Right: Comparison between the central slices of (a) the ground-truth, (b) the reconstruction from the true orientations, (c) the reconstruction obtained with our joint refinement method, (d) the reconstruction from the unrefined initial orientations.} 
    \label{fig:intro_pic}
\end{figure}

Single-particle cryo-electron microscopy (cryo-EM) aims at obtaining the three-dimensional (3D) atomic structures of biological macromolecules such as proteins or viruses. Replicates of a molecule of interest, in unknown orientations, are first imaged at cryogenic temperatures. From those 2D projections (Figure~\ref{fig:intro_pic} left), one then reconstructs the 3D density map of the molecule (Figure~\ref{fig:intro_pic} right), a computational process named ``single-particle analysis'' (SPA). The reconstruction task in SPA is extremely challenging due to the lack of knowledge on the projection directions, heavy noise and the blurring inherent with the point spread function (PSF) of the microscope. To tackle this difficulty, most methods start by estimating an \textit{ab-initio} model from class-averaged particle images. Then, this initial model is refined iteratively until a high-resolution map is obtained, a task named ``3D refinement''. 

\subsection{Standard 3D Refinement Techniques}

Currently, state-of-the-art refinement techniques~\textcolor{black}{\cite{Scheres2012,punjani2017cryosparc,Xmipp2013,Grigorieff2007}} produce a high-resolution density map by alternating between
\begin{enumerate}
    \item the reconstruction of the 3D density map for a given set of (however inaccurate) projection orientations; 
    \item the refinement of the projection orientations for all 2D particles based on the previously reconstructed 3D volume. 
\end{enumerate}

The reconstruction problem can be solved using different approaches such as algebraic methods~\cite{Gordon1970, Gilbert1972}, weighted back-projection (WBP)~\cite{Radermacher2006}, direct Fourier methods~\cite{Penczek2004, Abrishami2015, Carazo2015}, and iterative regularized approaches~\cite{Nilchian2013, Jaitly2010, donati2018fast}. 

In most SPA packages, direct Fourier methods based on the central-slice theorem are used. Those methods work adequately when the projections are sufficiently numerous and their speed is a key advantage. Unfortunately, their use is less appropriate in the presence of heavy noise, few projection measurements, or inaccurately known projection angles. 

The past years have seen \textcolor{black}{the appearance} of more robust iterative schemes that formulate the 3D reconstruction problem as a regularized optimization problem and enable the incorporation of prior knowledge on the underlying signal~\cite{sorzano2017survey}. Their downside is that iterative schemes usually come with a prohibitive computational cost if not carefully engineered. 

Several works have considerably improved this situation by proposing methods with increased speed. In particular, an important breakthrough came when a costly step of many reconstruction algorithms was shown to be quickly computable as a discrete convolution~\cite{delaney1996fast,vonesch2011fast,donati2018fast}. 

For the angular-refinement task, the most commonly used method is projection-matching~\textcolor{black}{~\cite{penczek1994ribosome, Baker1996}}. It compares every projection image against a finite set of clean templates obtained from the current estimate of the 3D density map and then assigns the angular parameters based on the closest match~\cite{Heymann2007}. Projection-matching hence performs angular assignment on a discretized orientation space. As a consequence, the quality of the angular refinement depends on the fineness of the discretization and the quality of the density map used to generate the templates. One bottleneck is that a fine discretization comes at the cost of large set of templates, which leads to a computationally expensive procedure. Moreover, methods based on projection-matching were found to degrade significantly in low SNR regimes~\cite{Sigworth2015} or when errors occur in the estimation of the density map used for the generation of the clean templates \cite{Punjani2017}.

Examples of joint-reconstruction methods that address 3D \textit{ab initio} modeling are found in~\cite{Barnett2017}, \cite{yves2018}. In~\cite{Barnett2017}, a frequency-marching approach that increases the resolution of the reconstruction is proposed. This leads to a smaller computational overhead in projection-matching steps. In~\cite{yves2018}, the density map parameters are updated through gradient descent while the projection orientations are recovered through exhaustive search on an $SO(3)$ grid, followed by convex optimization. Although they provide efficient 3D \textit{ab initio} modeling, these methods still suffer from the shortcomings of projection-matching. 

Finally, a joint-reconstruction framework for 2D tomography with unknown projection orientations is proposed in~\cite{Cheik2017}. The problem is solved through simulated annealing, which strongly limits its applicability to 3D tomography due to its high computational cost.

\subsection{Maximum-Likelihood Methods}
\label{sec:ML methods}
Scheres followed a Bayesian approach in \cite{Scheres2012} to formulate the 3D refinement problem as a maximum marginalized \textit{a posterior} (MAP) estimation~\textcolor{black}{~\cite{scheres2012bayesian}} that is solved by expectation maximization~\cite{Dempster1977}. This method is less sensitive to the initial model and brings higher robustness in low SNR regimes. However, its high computational complexity limits its applicability. 

Punjani \textit{et al.} proposed a  computationally efficient framework in~\cite{Punjani2017}. They formulated the 3D refinement problem as a MAP estimation and solved it by stochastic average gradient descent. They also used importance sampling to further reduce the cost of computing the marginalized likelihood. 

An advantage of maximum-likelihood-based methods is that they do not limit a particle image to a unique angular class. This leads to increased robustness in high-noise regimes compared to projection-matching procedures. However, they still involve some form of discretization of the projection orientations. In particular, they necessitate an overly fine discretization of the 3D orientation space, as well as a compactly supported grid over $\mathbb{R}^2$ for in-plane translations. Moreover, the marginalization process is usually computationally expensive. 

\subsection{Contributions}

In this work, we present a angular-refinement method for single-particle cryo-EM that jointly recovers the 3D density map and the orientation of each projection. This joint optimization problem is solved by letting the alternating-direction method of multipliers (ADMM) and gradient-descent steps take turns to update the density map and the orientations, respectively. 

We use an explicit derivation of the gradient of the objective function (Theorem~\ref{th:grad}) to optimize the orientations over a continuous space. Hence, a key advantage of the proposed approach over usual methods is that it avoids one to resort to a fine discretization of $SO(3)$ for the orientations and $\mathbb{R}^2$ for the in-plane translations.  Moreover, the computationally expensive step of projection-matching is skipped. 

By using fast algorithms, we are able to efficiently refine 3D density maps from sets of projections with poor initial angular estimation. We illustrate in Figure~\ref{fig:intro_pic} the type of refinements obtained with our joint-optimization framework, compared to a few baselines.

The paper is organized as follows: We describe in Section~\ref{sec:imaging_model} the image-formation model. In Section~\ref{sec:methods}, we detail our joint-optimization framework. The experimental setup is described in Section~\ref{sec:exp_conditions} and results are presented in Section~\ref{sec:results}. Finally, we conclude this work in Section \ref{sec:conclusion}. 


\subsection{Notations}

Sequences from $\mathbb{Z}^d \rightarrow \mathbb{R}$ are denoted by $\mathrm{c}[\cdot]$. Then, sequence samples are $\mathrm{c}[\mathbf{k}]$ with $\mathbf{k}=(k_1,\dots,k_d)\in\mathbb{Z}^d$. Bold lowercase letters (\textit{e.g.}, $\mathbf{c}$) represent vectors while bold uppercase letters are reserved for matrices (\textit{e.g.}, $\mathbf{H}$). All vectors are assumed to be column vectors unless otherwise stated. The $\ell_1$ and $\ell_2$ norms of the vector $\mathbf{c}\in\mathbb{R}^N$ are defined as $\lVert \mathbf{c} \rVert_1 \coloneqq \sum_{n=1}^N\abs{\mathrm{c}_n}$ and $\lVert \mathbf{c} \rVert_2 \coloneqq \big(\sum_{n=1}^N\abs{\mathrm{c}_n}^2 \big)^\frac{1}{2}$, respectively. The spaces $\ell_2(\mathbb{Z}^d)$ and $L_2(\mathbb{R}^d)$ contain finite-energy sequences and functions, respectively. The proximal operator of a convex functional $\mathcal{R}: \mathbb{R}^N \rightarrow \mathbb{R}$ is defined as $\text{prox}_{\mathcal{R}}(\mathbf{z};\mu) \coloneqq \,\argmin_{\mathbf{s}}\left( \frac{1}{2} \lVert \mathbf{s}-\mathbf{z} \rVert_2^2 + \mu \mathcal{R}(\mathbf{s}) \right)$, with $\mu\in\mathbb{R}$. The Fourier transform of $f$ is $\widehat f$. \textcolor{black}{The reflection of a function $f$ is denoted $f^\vee = f(-\cdot)$.}
Finally, the projection orientations and the in-plane translations are referred to as ``latent variables''.  

\section{Cryo-EM Imaging Model}
\label{sec:imaging_model}

\vspace{0.25cm}
\subsection{Imaging Model for a Single Orientation}

Let $V \in {L}_2(\mathbb{R}^3)$ denote the 3D density map of a molecule and let $\Omega_{\mathrm{2D}} \subset \mathbb{Z}^2$  be the discretized projection domain (see Figure~\ref{fig:angles}). The number of elements in $\Omega_\mathrm{2D}$ is $M = \sharp\Omega_\mathrm{2D}$.

We model a cryo-EM projection image $g: \Omega_\mathrm{2D} \rightarrow \mathbb{R}$ for an orientation $\bm{\uptheta} =(\theta_1,\theta_2,\theta_3) \in  [0,2\pi)\times [0,\pi]\times [0,2\pi) $ and an in-plane translation $\mathbf{t} = (t_\mathrm{1},t_\mathrm{2}) \in \mathbb{R}^2$ as
\begin{align}
g[\mathbf{m}]= (h * \mathcal{P}_{\bm{\uptheta}} (V) )(\bm{\Lambda}\mathbf{m}-\mathbf{t}) + \rm{\varepsilon}[\mathbf{m}],  \; 
\label{eq:image_form}
\end{align}
where $\bm{\Lambda} = \mathbf{diag}(\Delta_{\mathrm{1}},\Delta_{\mathrm{2}})$ is a diagonal matrix formed out of the sampling steps $\Delta_{\mathrm{1}}$ and $\Delta_{\mathrm{2}}$ of the projection domain and ${\varepsilon}: \Omega_\mathrm{2D} \rightarrow \mathbb{R}$ is an additive Gaussian white noise with zero mean and $\sigma^2$ variance. The operator $\mathcal{P}_{\bm{\uptheta}}:{L}_2(\mathbb{R}^3) \rightarrow {L}_2(\mathbb{R}^2)$ is the projection operator for the orientation $\bm{\uptheta}$ and $h \in {L}_2(\mathbb{R}^2)$ corresponds to the PSF of the microscope. The vectorization of $g$ is $\mathbf{g} = (g[\mathbf{m}])_{\mathbf{m} \in \Omega_{2D}}$ so that $\mathbf{g} \in \mathbb{R}^M$. The same goes for  $\varepsilon$  and $\bm{\varepsilon}$. 

\tdplotsetmaincoords{60}{110}
\pgfmathsetmacro{\rvec}{.8}
\pgfmathsetmacro{\thetavec}{30}
\pgfmathsetmacro{\phivec}{60}
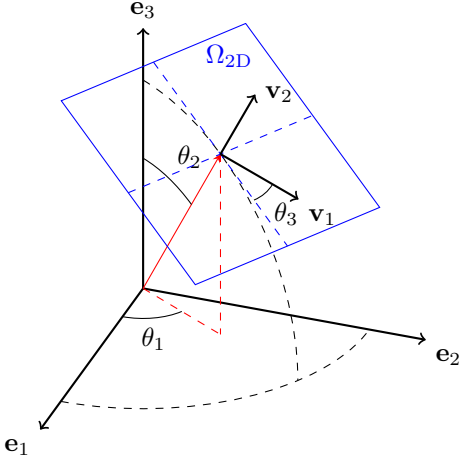
\begin{figure}[t!]
\centering
\begin{tikzpicture}[scale=4,tdplot_main_coords]
    \coordinate (O) at (0,0,0);
    \draw[thick,->] (0,0,0) -- (1,0,0) node[anchor=north east]{$\mathbf{e}_1$};
    \draw[thick,->] (0,0,0) -- (0,1,0) node[anchor=north west]{$\mathbf{e}_2$};
    \draw[thick,->] (0,0,0) -- (0,0,1) node[anchor=south]{$\mathbf{e}_3$};
    \tdplotsetcoord{P}{\rvec}{\thetavec}{\phivec}
    \draw[-stealth,color=red] (O) -- (P);
    \draw[dashed, color=red] (O) -- (Pxy);
    \draw[dashed, color=red] (P) -- (Pxy);
    \tdplotdrawarc{(O)}{0.2}{0}{\phivec}{anchor=north}{$\theta_1$}
    \tdplotsetthetaplanecoords{\phivec}
    \tdplotdrawarc[tdplot_rotated_coords]{(0,0,0)}{0.5}{0}%
        {\thetavec}{anchor=south west}{$\theta_2$}
    \draw[dashed,tdplot_rotated_coords] (\rvec,0,0) arc (0:90:\rvec);
    \draw[dashed] (\rvec,0,0) arc (0:90:\rvec);
    \tdplotsetrotatedcoords{\phivec}{\thetavec}{0}
    \tdplotsetrotatedcoordsorigin{(P)}
    \draw[dashed,blue,tdplot_rotated_coords,-] (-.4,0,0)
        -- (.4,0,0) node[anchor=north west]{};
    \draw[dashed,blue,tdplot_rotated_coords,-] (0,-.4,0)
        -- (0,.4,0) node[anchor=west]{};
    \draw[blue,tdplot_rotated_coords,-]  (-.4,.4,0) -- (.4,.4,0)  -- (.4,-.4,0) -- (-.4,-.4,0) -- (-.4,.4,0)   node[anchor=north]{};
    \tdplotdrawarc[tdplot_rotated_coords]{(0,0,0)}{0.2}{0}%
        {30}{anchor=north west,color=black}{$\theta_3$}
    \tdplotsetrotatedcoords{\phivec}{\thetavec}{30}
    \draw[thick,tdplot_rotated_coords,->] (0,0,0)
        -- (.3,0,0) node[anchor=north west]{$\mathbf{v}_1$};
    \draw[thick,tdplot_rotated_coords,->] (0,0,0)
        -- (0,.3,0) node[anchor=west]{$\mathbf{v}_2$};
    \node[blue] at (0.4,0.45,1.2) {$\Omega_{\mathrm{2D}}$};    
    \tdplotsetrotatedthetaplanecoords{45}
\end{tikzpicture}
\caption{3D Geometry of the imaging model for an orientation~$\boldsymbol{\uptheta}=(\theta_1,\theta_2,\theta_3)$. The Euler angles $\theta_1$, $\theta_2$, and $\theta_3$ represent the rotation, the tilt, and the in-plane rotation in the projection plane, respectively.}
\label{fig:angles}
\end{figure}

\subsection{Discretization}

To discretize the 3D density map $V$, we follow a generalized sampling scheme~\cite{Unser2000} and define
\begin{equation}
V(\mathbf{x}) = \sum\limits_{\mathbf{k} \in \mathbb{Z}^3} {c}[\mathbf{k}] \varphi({\mathbf{x}}-\mathbf{k}), \; \forall \mathbf{x} \in \mathbb{R}^3,\label{eq:sampling}
\end{equation}
where $\varphi \in {L}_2(\mathbb{R}^3)$ is a given basis function and $c[\cdot] \in \ell_2(\mathbb{Z}^3)$ is a sequence that contains the coefficients of $V$ in the reconstructing space. The sampling step in the object domain is assumed to be equal to one, without loss of generality.

In this work, we choose $\varphi$ to be the optimized Kaiser-Bessel window function (KBWF)~\cite{Nilchian2015}
    \begin{equation}\label{eq:KBWF}
       \mkern-12mu  \varphi(\mathbf{x}) \mkern-2mu  =  \mkern-2mu \left\lbrace \mkern-10mu
        \begin{array}{ll}
              \frac{\beta_a(\|\mathbf{x}\|)^m  I_m\left(\alpha \beta_a(\|\mathbf{x}\|) \right)}{I_m(\alpha)},   \mkern-10mu &  \|\mathbf{x}\| \in [0,a] \\
           0,  & \text{otherwise} ,
        \end{array} \right. 
    \end{equation}
where $\beta_a(r)= \sqrt{1 - \left({r}/{a}\right)^2} $, $a>0$ is the support radius, $\alpha>0$ the window taper, and $I_m$ the modified Bessel function of order $m$.
KBWFs are well suited for tomographic reconstruction in reason of their isotropy and compact support~\cite{Nilchian2015, donati2018fast}. Moreover, the x-ray transform of KBWF does not depend on the orientation $\bm{\uptheta}$ and admits a closed-form expression~\cite{Lewitt1990}. It was shown in~\cite{Nilchian2015} that a KBWF represents functions very effectively when using specific parameter values (\textit{e.g.},  $m=2$, $a=4$, and $\alpha=19$).

Because the density map $V$ is compactly supported, the sequence  $c[\cdot] \in \ell_2(\mathbb{Z}^3)$ can be restricted to a finite number of nonzero coefficients $\mathbf{c}=({c}[\mathbf{k}])_{\mathbf{k}\in \Omega_{\mathrm{3D}}}$, where $\Omega_{\mathrm{3D}} \subset \mathbb{Z}^3$ and $N = \sharp\Omega_\mathrm{3D}$. 

We then substitute \eqref{eq:sampling} in \eqref{eq:image_form}, and use the linearity and the pseudo-translation invariance of the x-ray transform~\cite{Natterer2001} to obtain a discrete version of the forward model, as in
\begin{align}
\mkern-5mu {g}[\mathbf{m}] & = \mkern-12mu \sum\limits_{\mathbf{k} \in \Omega_{\mathrm{3D}}}  \mkern-9mu  {c}[\mathbf{k}] (h * \mathcal{P}_{\bm{\uptheta}} (\varphi) )(\bm{\Lambda}\mathbf{m} - \mathbf{M}_{\bm{\uptheta}^\perp} \mathbf{k} - \mathbf{t}) + \varepsilon[\mathbf{m}]. \label{eq:disc_forward}
\end{align}
Here, $\mathbf{M}_{\bm{\uptheta}^{\perp}} \in \mathbb{R}^{2 \times 3}$ is the orthogonal projector operator
\begin{equation}
\mathbf{M}_{\bm{\uptheta}^{\perp}} \mkern-5mu = \mkern-5mu   \begin{pmatrix} C_1C_2C_3 - S_1S_3 & \mkern-5mu C_3S_1+C_1C_2S_3 & \mkern-5mu-C_1S_2 \\ 
-C_1S_3-C_2C_3S_1 & \mkern-5muC_1C_3 - C_2S_1S_3 & \mkern-5mu S_1S_2 \end{pmatrix}, \label{eq:M_omega}
\end{equation}
where $\forall i = \{1,2, 3\}, \, C_i = \cos(\theta_i)$ and $S_i=\sin(\theta_i)$.

Finally, we write \eqref{eq:disc_forward} as
\begin{equation}
    \mathbf{g} = \mathbf{H}(\bm{\uptheta},\mathbf{t}) \, \mathbf{c}+ \mathrm{\bm{\upvarepsilon}}, \label{eq:linear_proj_model}
\end{equation}
where $\mathbf{H}(\bm{\uptheta},\mathbf{t}) \in \mathbb{R}^{M \times  N}$ is the discrete imaging operator for orientation $\bm{\uptheta}$ and in-plane translation $\mathbf{t}$.

\subsection{Global Imaging Model}

We now consider a set of $P$ projection images (indexed as $\mathbf{g}_p$) such that $\mathbf{g}=\{\mathbf{g}_p \in \mathbb{R}^{M} \}_{p =1}^P$. Similarly, the set of projection orientations is defined as $\boldsymbol{\Theta} = \{\bm{\uptheta}_p\in \Theta \}_{p =1}^P$ and the set of in-plane translations as $\boldsymbol{\Gamma}=\{\mathbf{t}_p\in \mathbb{R}^2 \}_{p =1}^P$. 

The global imaging model is thus given by
\begin{align}
\label{equ:global-imaging-model}
    \mathbf{g} = \mathbf{H}(\boldsymbol{\Theta}, \boldsymbol{\Gamma}) \, \mathbf{c}+ \bm{\upvarepsilon},
\end{align}
where \begin{equation}
   \mathbf{g} = \begin{bmatrix}
                \mathbf{g}_1 \\ \vdots \\ \mathbf{g}_P
                \end{bmatrix}, \; 
    \mathbf{H}(\boldsymbol{\Theta}, \boldsymbol{\Gamma}) =   
    \begin{bmatrix}
                \mathbf{H}(\bm{\uptheta}_1,\mathbf{t}_1) \\ \vdots \\ \mathbf{H}(\bm{\uptheta}_P,\mathbf{t}_P)
                \end{bmatrix}, 
    \bm{\upvarepsilon} = \begin{bmatrix}
                \bm{\upvarepsilon}_1 \\ \vdots \\ \bm{\upvarepsilon}_P
                \end{bmatrix}
\end{equation}

For the sake of clarity we shall thereafter use the notations  $\mathbf{H}^T\mathbf{H}(\boldsymbol{\Theta}, \boldsymbol{\Gamma})=(\mathbf{H}(\boldsymbol{\Theta}, \boldsymbol{\Gamma}))^T\mathbf{H}(\boldsymbol{\Theta}, \boldsymbol{\Gamma})$ and $\mathbf{H}^T(\boldsymbol{\Theta}, \boldsymbol{\Gamma}) = (\mathbf{H}(\boldsymbol{\Theta}, \boldsymbol{\Gamma}))^T$.

\section{Joint Angular Refinement and Reconstruction}
\label{sec:methods}


\subsection{Joint-Optimization Framework}

Our goal is to jointly estimate the unknown variables in~\eqref{equ:global-imaging-model}, which are the coefficients~$\mathbf{c}$ of the density map, the projection orientations $\boldsymbol{\Theta}$, and the in-plane translations $\boldsymbol{\Gamma}$. To do so, we express the refinement procedure as a regularized least-squares minimization
\begin{equation}
   \big(\hat{\mathbf{c}}, \hat{\boldsymbol{\Theta}}, \hat{\boldsymbol{\Gamma}}\big) \in \left\lbrace \mathrm{arg} \, \underset{\mathbf{c},\boldsymbol{\Theta}, \boldsymbol{\Gamma}}{\mathrm{min}}  \;    \mathcal{J}\big(\mathbf{c}, \boldsymbol{\Theta}, \boldsymbol{\Gamma}\big) \right\rbrace, \label{eq:opt}
\end{equation}
where 
\begin{equation}
    \mathcal{J}\big(\mathbf{c}, \boldsymbol{\Theta}, \boldsymbol{\Gamma}\big) = \frac{1}{2} \Vert \mathbf{g} - \mathbf{H}(\boldsymbol{\Theta}, \boldsymbol{\Gamma}) \, \mathbf{c}\Vert_2^2+ \lambda \mathcal{R}( \mathbf{L} \mathbf{c}).
    \label{eq:funcOpt}
\end{equation}
Here, $\mathcal{R}$ is a sparsity-promoting functional and $\mathbf{L}$ a linear operator. Together, they are used to inject prior knowledge into the reconstruction process. As an example, setting $\mathbf{L}=\bm{\nabla}$ (i.e., the gradient operator) and $\mathcal{R}=\|\cdot\|_{2,1}$ leads to the popular total-variation (TV) regularization~\cite{rudin1992nonlinear}. The regularization parameter $\lambda$ controls the balance between the data-fidelity and the regularization terms. 

To solve \eqref{eq:opt}, we alternate between the minimization over $\mathbf{c}$ and the minimization over $\boldsymbol{\Theta}$ and $\boldsymbol{\Gamma}$. Although the objective function in \eqref{eq:opt} is convex with respect to $\mathbf{c}$, it is not convex with respect to the latent variables $\boldsymbol{\Theta}$ and $\boldsymbol{\Gamma}$. Moreover, it is smooth with respect to $\boldsymbol{\Theta}$ and $\boldsymbol{\Gamma}$, but usually not smooth (due to $\mathcal{R}$) with respect to $\mathbf{c}$. This dictates the choice of two different minimization procedures within the proposed alternating scheme. 

For the minimization of $\mathcal{J}$ with respect to $\mathbf{c}$, we use ADMM~\cite{Boyd2011}, which allows us to split the problem into a sequence of simpler subproblems (see Section~\ref{sec:update_c}). Then, taking benefit from the differentiability of $\mathcal{J}$ with respect to $\boldsymbol{\Theta}$ and $\boldsymbol{\Gamma}$, the latent variables are updated using gradient-descent with line-search (see Section~\ref{sec:update_latent}). The outline of this joint optimization procedure is given in Algorithm~\ref{alg:refinement} and is implemented within the framework of the GlobalBioIm library\footnote{\texttt{http://bigwww.epfl.ch/algorithms/globalbioim/}}~\cite{soubies2018pocket}.

Note that, at Line~\ref{line:ADMM} of  Algorithm~\ref{alg:refinement}, we use the notation $\mathrm{ADMM}\left(\mathcal{J}( \cdot,\boldsymbol{\Theta}^k, \boldsymbol{\Gamma}^k), \mathbf{c}^{k} \right)$  to refer to the minimization of $\mathcal{J}( \cdot,\boldsymbol{\Theta}^k, \boldsymbol{\Gamma}^k)$ using ADMM initialized with~$\mathbf{c}^{k}$. We do the same for the gradient-descent algorithm (see Line~\ref{line:GD}). 
 
\begin{algorithm}[t]
  \caption{Joint-Optimization Framework
}
  \label{alg:refinement}
  \textbf{Require:} $\mathbf{c}^0$,$\bm{\Theta}^0$,$\bm{\Gamma}^0$
  \begin{algorithmic}[1]
    \State $k=0$
    \While {not converged} 
    \LineComment{Update the density map:}
    \State 
    $\mathbf{c}^{k+1} = \mathrm{ADMM}\left(\mathcal{J}( \cdot,\boldsymbol{\Theta}^k, \boldsymbol{\Gamma}^k), \mathbf{c}^{k} \right)$ \label{line:ADMM}
     \LineComment{Update the latent variables:} 
      \State 
      $ (\bm{\Theta}^{k+1},\bm{\Gamma}^{k+1}) =  \mathrm{GD}\left(\mathcal{J}( \mathbf{c}^{k+1},\cdot,\cdot), \bm{\Theta}^k,\mathbf{\Gamma}^k \right)$ \label{line:GD}
        \State $ k \gets k+1$
        \EndWhile
  \end{algorithmic}
   \textbf{Return:} $ \mathbf{c}^{k},\bm{\Theta}^{k}, \mathbf{\Gamma}^{k}$
\end{algorithm}


\subsection{Update of the Density Map} \label{sec:update_c}

Given $\boldsymbol{\Theta}$ and $\boldsymbol{\Gamma}$, the reconstruction task consists in solving 
\begin{equation}
    \label{funcOpt-min-wrt-c}
    \hat{\mathbf{c}}= \textrm{arg} \min_{\mathbf{c}} \; \mathcal{J}(\mathbf{c},\boldsymbol{\Theta}, \boldsymbol{\Gamma}).
\end{equation}
To do so, we use the ADMM scheme proposed in~\cite{donati2018fast}. The core idea is to split~\eqref{funcOpt-min-wrt-c} by introducing an auxiliary variable $\mathbf{u}$ so that
\begin{align}
\label{eq:admm-minization-constrained}
\begin{split}
\hat{\mathbf{c}}= \textrm{arg} \min_{\mathbf{c}} & \; \left(\frac{1}{2} \Vert \mathbf{g} - \mathbf{H}(\boldsymbol{\Theta}, \boldsymbol{\Gamma}) \, \mathbf{c}\Vert_2^2 +\lambda \mathcal{R}(\mathbf{u}) \right) \\
 s.t. & \quad \mathbf{u}=\mathbf{L}\mathbf{c}.
\end{split}
\end{align} 
Then, the ADMM algorithm alternates between three steps, as summarized in Algorithm~\ref{alg:volume_update}. 

When TV regularization is used, the proximal operator at Line~\ref{algo:prox-u} admits a closed-form expression that can be computed efficiently~\cite{combettes2008proximal}. Then, the linear step at Line~\ref{algo:standard-admm-linear-step} is solved iteratively using a conjugate-gradient algorithm together with a fast formulation of the $\mathbf{H}^T\mathbf{H}(\boldsymbol{\Theta}, \boldsymbol{\Gamma})$ term~\cite{donati2018fast}. Finally, Line~\ref{admmcg-update-u} corresponds to a simple update of the dual variable $\widetilde{\mathbf{u}}$, while $\rho>0$ is a penalty parameter.

For the sake of completeness, the full set of equations behind the reconstruction algorithm is provided in Appendix~\ref{subsec:appendix-ADMM}.

\begin{algorithm}[t]
\caption{ADMM (Update of the Density Map)}
\label{alg:volume_update}
\vspace{0.1cm}
\textbf{Require:} $\mathbf{g}$, $\boldsymbol{\Theta}$, $\boldsymbol{\Gamma}$, $\mathbf{c}^0$, $\rho>0$ 
\begin{algorithmic}[1]
\State $\mathbf{u}^0= \mathbf{Lc}^0$, $\tilde{\mathbf{u}}^0= \mathbf{u}^0$
\State $k=0$ 
\While {$k < K_{\tiny \mathrm{ADMM}}$}
\State $\mathbf{u}^{k+1} = \text{prox}_{\mathcal{R}}\big(\mathbf{L}\mathbf{c}^k-{\widetilde{\mathbf{u}}^k}/{\rho}; \, {\lambda}/{\rho}\big)$ \label{algo:prox-u} 
\State $\mathbf{b} = (\mathbf{H}(\boldsymbol{\Theta}, \boldsymbol{\Gamma}))^T \mathbf{g} + \rho \mathbf{L}^T (\mathbf{u}^{k+1} - \tilde{\mathbf{u}}^k/\rho)$
\State $\mathbf{c}^{k+1} = \left(\mathbf{H}^T\mathbf{H}(\boldsymbol{\Theta}, \boldsymbol{\Gamma}) + \rho \mathbf{L}^T \mathbf{L} \right)^{-1} \mathbf{b}$ \label{algo:standard-admm-linear-step}
\State $\widetilde{\mathbf{u}}^{k+1} = \widetilde{\mathbf{u}}^{k}+\rho\big(\mathbf{u}^{k+1}-\mathbf{L}\mathbf{c}^{k+1}\big)$ \label{admmcg-update-u} 
\State $k\gets k+1$ 
\EndWhile
\end{algorithmic}
\textbf{Return:} $\mathbf{c}^{ K_{\tiny \mathrm{ADMM}}}$
\end{algorithm}


\subsection{Update of the Latent Variables} \label{sec:update_latent}

   Let us first remark that the least-squares term in~\eqref{eq:funcOpt} can be written as
   \begin{equation}
        \frac{1}{2} \Vert \mathbf{g} - \mathbf{H}(\boldsymbol{\Theta}, \boldsymbol{\Gamma}) \, \mathbf{c}\Vert_2^2 = \frac{1}{2} \sum_{p=1}^P \|\mathbf{g}_p - \mathbf{H}(\bm{\uptheta}_p,\mathbf{t}_p) \, \mathbf{c} \|_2^2.
   \end{equation}
   Hence, when $\mathbf{c}$ is fixed, the minimization of $\mathcal{J}(\mathbf{c},\cdot,\cdot)$ amounts to solve
   \begin{equation}
       (\hat{\bm{\uptheta}}_p, \hat{\mathbf{t}}_p)  \in \left\lbrace    \mathrm{arg} \min_{\bm{\uptheta}, \mathbf{t}} \;\mathcal{J}_p(\bm{\uptheta}, \mathbf{t}) \right\rbrace \label{eq:latentVarOpt}
   \end{equation}
    for all $ p \in \{1,\ldots,P\}$, where  $\mathcal{J}_p : (\bm{\uptheta}, \mathbf{t}) \mapsto \mathbb{R}$ is defined as
   \begin{equation}
        \mathcal{J}_p(\bm{\uptheta},\mathbf{t}) = \frac{1}{2} \Vert \mathbf{g}_p - \mathbf{H}(\bm{\uptheta}, \mathbf{t}) \, \mathbf{c}\Vert_2^2.
        \label{eq:latentVarOpt-objFunction}
   \end{equation}
   As the objective function $\mathcal{J}_p$ is differentiable, the minimization in~\eqref{eq:latentVarOpt} can be achieved using gradient-descent steps. 
   {\color{black}
   Hence, we first need to compute the gradients
   \begin{align}
       & \bm{\nabla}_{\bm{\uptheta}}  \mathcal{J}_p(\bm{\uptheta}, \mathbf{t}) =  \left( \frac{ \partial \mathcal{J}_p}{\partial \theta_1}(\bm{\uptheta}, \mathbf{t}) , \frac{ \partial \mathcal{J}_p}{\partial \theta_2}(\bm{\uptheta}, \mathbf{t}) , \frac{ \partial \mathcal{J}_p}{\partial \theta_3}(\bm{\uptheta}, \mathbf{t})  \right) \label{eq:PartialTheta} \\
       & \bm{\nabla}_{\mathbf{t}}\mathcal{J}_p(\bm{\uptheta},\mathbf{t}) = \left( \frac{ \partial \mathcal{J}_p}{\partial t_1}(\bm{\uptheta}, \mathbf{t}), \frac{ \partial \mathcal{J}_p}{\partial t_2}(\bm{\uptheta}, \mathbf{t}) \right). \label{eq:PartialT}
   \end{align}
   

   The explicit expressions of these  quantities are provided in Theorem~\ref{th:grad}. 
   \begin{theorem}\label{th:grad} Let $\varphi$ be an isotropic kernel and $\mathbf{H}(\bm{\uptheta},\mathbf{t}) \in \mathbb{R}^{M \times N}$ be defined by~\eqref{eq:disc_forward}.  Then, for $v \in \{\theta_1,\theta_2,\theta_3,t_1,t_2\}$, there exists $\mathbf{r}_v \in \mathbb{R}^N$ and $\mathbf{q}_v \in \mathbb{R}^N$ such that
   \begin{equation}\label{eq:thGrad}
       \frac{ \partial \mathcal{J}_p}{\partial v}(\bm{\uptheta}, \mathbf{t}) = \frac12 \mathbf{c}^T \left( \mathbf{r}_v \ast \mathbf{c} - 2 \mathbf{q}_v \right).
   \end{equation}
    Moreover, $\forall \mathbf{k} \in \Omega_\mathrm{3D}$,\\
        $\bullet$ if $v = \theta_i$ for $i \in \{1,2,3\}$,
         \begin{align}
                &{r}_v[\mathbf{k}] =  \frac{1}{\mathrm{det}(\bm{\Lambda})} \left( \frac{\partial \mathbf{M}_{{\bm{\uptheta}}^{\perp}} }{\partial \theta_i}   \mathbf{k} \right)^{T} \mkern-10mu \bm{\nabla} \left( \psi \ast \psi^\vee \right) (\mathbf{M}_{\bm{\uptheta}^{\perp}} \mathbf{k}), \\
                 \mkern-8mu & {q}_v[\mathbf{k}] = \frac{1}{\mathrm{det}(\bm{\Lambda})}  \bigg( \frac{\partial \mathbf{M}_{{\bm{\uptheta}}^{\perp}} }{\partial \theta_i}   \mathbf{k} \bigg)^{\mkern-3mu T} \mkern-5mu  \bm{\nabla}(g_p \ast \psi^\vee)  (\mathbf{M}_{\bm{\uptheta}^{\perp}} \mathbf{k}+\mathbf{t}),
        \end{align}
        $\bullet$ if $v = t_j$ for $j \in \{1,2\}$,
        \begin{align}
             &{r}_v[\mathbf{k}] = 0, \\
            &{q}_v[\mathbf{k}] = \frac{1}{\mathrm{det}(\bm{\Lambda})}   \frac{\partial (g_p \ast \psi^\vee)}{\partial y_j} (\mathbf{M}_{\bm{\uptheta}^{\perp}} \mathbf{k} + \mathbf{t}) ,
        \end{align}
    where  $\psi : \mathbf{y}=(y_1,y_2) \mapsto (h * \mathcal{P}(\varphi))(\mathbf{y})$, $\frac{\partial  \mathbf{M}_{{\bm{\uptheta}}^{\perp}}}{\partial \theta_i} \in \mathbb{R}^{2 \times 3}$ contains the entry-wise derivatives of $\mathbf{M}_{{\bm{\uptheta}}^{\perp}}$, and $g_p$ denotes the continuous counterpart of $\mathbf{g}_p$ (\ie interpolated values). 
   \end{theorem}

The proof is given in Appendix~\ref{sec:ProofGrad} and includes details on the gradients of $\psi \ast \psi^\vee$ and $g_p \ast \psi^\vee$. In particular, we show that they depend on $\mathcal{P}(\varphi)$ and $ \frac{\partial \mathcal{P}(\varphi) }{\partial y_j}$ whose expressions are provided in Proposition~\ref{propo:KBWF} for the specific case of the KBWF $\varphi$ in~\eqref{eq:KBWF}.  
   }
   
    \begin{proposition} \label{propo:KBWF} For the KBWF $\varphi$ given in~\eqref{eq:KBWF}, we have
    \begin{align}
       & \mkern-8mu \mathcal{P}(\varphi)(\mathbf{y}) = aA \, \beta_a(\|\mathbf{y}\|)^{m+\frac12}  I_{m + \frac12} \mkern-5mu\left( \alpha \beta_a(\|\mathbf{y}\|) \right),  \\
        &  \mkern-8mu \frac{\partial \mathcal{P}(\varphi) }{\partial y_v}(\mathbf{y}) =  \mkern-5mu - \frac{\alpha y_vA }{a}\beta_a(\|\mathbf{y}\|)^{m-\frac12}  I_{m-\frac12}(\alpha \beta_a(\|\mathbf{y}\|)),
    \end{align}
    where $A=\frac{\sqrt{2\pi/\alpha}}{I_m(\alpha)}$.
    \end{proposition}
    
The proof is given in Appendix~\ref{sec:ProofKBWF}.  

Equipped with those gradient expressions, we deploy a semi-coordinate-wise gradient-descent to solve~\eqref{eq:latentVarOpt}, as summarized in Algorithm \ref{alg:angular_assignment}. At each iteration, the parameters $\bm{\uptheta}$ and $\mathbf{t}$ are updated sequentially, which allows for the use of different stepsizes between orientation and in-plane translation. This is crucial to account for the different dynamics between these two variables. Moreover, we use adaptive steps that are selected according to a backtracking line-search method~\cite{armijo1966,Nocedal2006}. Given  an initial value, the step is decreased through the parameter $\eta \in (0,1)$ until the cost that corresponds to the updated variable is smaller than its current value (conditions checked in Steps \ref{line:AlgoGD-1} and \ref{line:AlgoGD-2}).

Finally, to further accelerate the update of the latent variables, we divide the projection set $\{1,\ldots,P\}$ into mini-batches and process them in parallel. It is the separability of the objective function in~\eqref{eq:latentVarOpt}, related to the independence of projection images, that makes this parallelization possible.

\begin{algorithm}[t!]
  \caption{GD (Update of the Latent Variables)}
  \label{alg:angular_assignment}
 \textbf{Require:} $\alpha^0_{\bm{\uptheta}}>0$, $\alpha^0_{\mathbf{t}}>0$, $\eta$ $\in (0,1)$, $\bm{\Theta}^0$, $\bm{\Gamma}^0$, $\mathbf{c}$
  \begin{algorithmic}[1]
    \NoDoFor{$p=1,\ldots,P$} 
    \State $k = 0$
    \NoDoWhile{$k < K_{\mathrm{GD}}$}
    \LineComment[2\dimexpr\algorithmicindent]{Update $\bm{\uptheta}_p$}
     \State $\alpha_{\bm{\uptheta}} \gets \alpha^0_{\bm{\uptheta}}$
    \State $\bm{\uptheta}^{k+1}_p = \bm{\uptheta}^{k}_p - \alpha_{\bm{\uptheta}} \bm{\nabla}_{\bm{\uptheta}} \mathcal{J}_p(\bm{\uptheta}_p^k,\mathbf{t}_p^k)$
    \NoDoWhile{$\mathcal{J}_p(\bm{\uptheta}_p^{k+1},\mathbf{t}_p^k) > \mathcal{J}_p(\bm{\uptheta}_p^k,\mathbf{t}_p^k)$} \label{line:AlgoGD-1}
    \State $\alpha_{\bm{\uptheta}} \leftarrow \eta \alpha_{\bm{\uptheta}}$
    \State $\bm{\uptheta}^{k+1}_p = \bm{\uptheta}^{k}_p - \alpha_{\bm{\uptheta}} \bm{\nabla}_{\bm{\uptheta}} \mathcal{J}_p(\bm{\uptheta}_p^k,\mathbf{t}_p^k)$
    \EndWhile
    \LineComment[2\dimexpr\algorithmicindent]{Update $\mathbf{t}_p$}
     \State $\alpha_{\mathbf{t}} \gets \alpha^0_{\mathbf{t}}$
    \State $\mathbf{t}^{k+1}_p = \mathbf{t}^{k}_p - \alpha_{\mathbf{t}} \bm{\nabla}_{\mathbf{t}} \mathcal{J}_p(\bm{\uptheta}_p^{k+1},\mathbf{t}_p^k)$
    \NoDoWhile{$\mathcal{J}_p(\bm{\uptheta}_p^{k+1},\mathbf{t}_p^{k+1}) > \mathcal{J}_p(\bm{\uptheta}_p^{k+1},\mathbf{t}_p^k)$}  \label{line:AlgoGD-2}
    \State $\alpha_{\mathbf{t}} \leftarrow  \eta \alpha_{\mathbf{t}}$
    \State $\mathbf{t}^{k+1}_p = \mathbf{t}^{k}_p - \alpha_{\mathbf{t}} \bm{\nabla}_{\mathbf{t}} \mathcal{J}_p(\bm{\uptheta}_p^{k+1},\mathbf{t}_p^k)$
    \EndWhile
    \State $k \gets k+1$
    \EndWhile
    \EndFor
  \end{algorithmic}
   \textbf{Return:} ${\bm{\Theta}}^{K_{\mathrm{GD}}},{\bm{\Gamma}}^{K_{\mathrm{GD}}}$
\end{algorithm}

{\color{black}
\subsection{Computational Complexity}
\label{sec:ComplexityComparaison}
We compare the computational complexity of the proposed latent variable update to that of projection matching.
Let $n$ and $m$ be such that $N = n^3$ and $M = m^2$ (\ie, $\mathbf{c} \in \mathbb{R}^{n \times n \times n}$ and $\mathbf{g}_p \in \mathbb{R}^{m \times m}$).

\subsubsection{Projection Matching}
Each iteration of projection matching consists of two steps.
\begin{itemize}
    \item \textit{Generation of Clean Templates.} Given the current density map $\mathbf{c}$, evaluate the forward model $\mathbf{H}(\bm{\uptheta},\mathbf{0}_{\mathbb{R}^2}) \, \mathbf{c}$ for $N_{\theta_1}N_{\theta_2}$ different values of $\bm{\uptheta}=(\theta_1,\theta_2,0)$ obtained by sampling $[0,2\pi)$ with $N_{\theta_1}$ points and $[0,\pi]$ with $N_{\theta_2}$ points. Denoting by $C_H$ the cost of one evaluation of the forward model, the computational complexity of this step is $O(N_{\theta_1}N_{\theta_2} C_H)$.
    \item \textit{Matching Projection Images $\mathbf{g}_p$.} Each projection image $\mathbf{g}_p$ ($p\in \{1,\ldots,P\}$) is compared against $N_{\theta_1}N_{\theta_2}$ clean templates. This requires rotation and in-plane translation alignment whose complexity is $O(m^2 \log m)$ if done efficiently using polar Fourier transform~\cite{Yang2008}, spherical harmonics~\cite{Lee2007}, or steerable basis functions~\cite{Zhao2014}.
\end{itemize}
The overall complexity of template matching is thus given by $O( N_{\theta_1} N_{\theta_2}( P m^2 \log m +  C_H))$. The cost of $C_H$ depends on its implementation. An efficient way to compute it can, for example, rely on the Fourier-slice theorem and the use of non-uniform FFT. This strategy roughly requires one 3D-FFT of the volume $\mathbf{c}$, one interpolation step to extract the central slice perpendicular  to  the  projection  direction, and one inverse 2D-FFT of this slice. This gives $C_H= n^3 \log n + m^2 + m^2 \log m$. With such an implementation, the overall complexity would thus be $O( N_{\theta_1} N_{\theta_2}( P m^2 \log m +  n^3 \log n))$.

\subsubsection{Proposed Update Scheme}
According to~\eqref{eq:thGrad} in Theorem~\ref{th:grad}, the evaluation of the partial derivative  $\partial \mathcal{J}_p / \partial v$ can be done at the cost of a 3D convolution (only required when $v=\theta_i$), a component-wise subtraction, and a scalar product. This gives a complexity of $O(n^3 \log n)$. To this has to be added the cost of computing $\mathbf{r}_v$ and $\mathbf{q}_v$ in Theorem~\ref{th:grad}. First, let us remark that $\frac{\partial  \mathbf{M}_{{\bm{\uptheta}}^{\perp}}}{\partial \theta_i}$ is known in closed form from~\eqref{eq:M_omega}. Hence, the complexity of computing $\frac{\partial  \mathbf{M}_{{\bm{\uptheta}}^{\perp}}}{\partial \theta_i} \mathbf{k}$ for all $\mathbf{k} \in \Omega_{\mathrm{3D}}$, is $O(n^3)$. Then, we distinguish two situations:
\begin{itemize}
    \item \textit{Explicit Expressions of $\bm{\nabla} \left( \psi \ast \psi^\vee \right) $ and}\footnote{Note that $\frac{\partial (g_p \ast \psi^\vee)}{\partial y_j}$ is nothing else than the $j$th component of $\bm{\nabla}(g_p \ast \psi^\vee)$.} \textit{$\bm{\nabla}(g_p \ast \psi^\vee)$ are Known}. Given $\bm{\uptheta}$ and $\mathbf{t}$, the computation of $\mathbf{r}_v$ and $\mathbf{q}_v$ amounts to their sampling at points $\mathbf{M}_{\bm{\uptheta}^{\perp}} \mathbf{k}$ (or $\mathbf{M}_{\bm{\uptheta}^{\perp}} \mathbf{k}+\mathbf{t}$), for $\mathbf{k} \in \Omega_{\mathrm{3D}}$, followed by a scalar product with $\frac{\partial  \mathbf{M}_{{\bm{\uptheta}}^{\perp}}}{\partial \theta_i} \mathbf{k}$, resulting in an overall complexity of~$O(n^3)$.
    \item \textit{Explicit Expressions of  $\bm{\nabla} \left( \psi \ast \psi^\vee \right) $ and $\bm{\nabla}(g_p \ast \psi^\vee)$ are Not Known}. Due to their independence upon $\bm{\uptheta}$ and $\mathbf{t}$, the relevant quantities can thus be evaluated \textit{once} (optionally upsampled) on the grid $\Omega_\mathrm{2D}$ using~\eqref{eq:GradPsiConvPsiVee} together with Proposition~\ref{propo:KBWF} and discrete convolutions (complexity of $O(m^2 \log m)$). Having this precomputed quantity saved as a lookup table, the evaluation of $\bm{\nabla} \left( \psi \ast \psi^\vee \right) $ and $\bm{\nabla}(g_p \ast \psi^\vee)$ at points  $\mathbf{M}_{\bm{\uptheta}^{\perp}} \mathbf{k}$ (or $\mathbf{M}_{\bm{\uptheta}^{\perp}} \mathbf{k}+\mathbf{t}$) is done by interpolation. Hence, here again, the computational complexity is $O(n^3)$.
\end{itemize}
Considering that there are~$P$ projection images and that $K_\mathrm{GD}$ iterations of gradient descent are performed at each update of the latent variables (see Algorithm~\ref{alg:refinement}), we obtain an overall complexity of $O(P K_\mathrm{GD} n^3 \log n)$. 

Finally, given that $K_\mathrm{GD}$ is typically small (for example in our experiments $K_\mathrm{GD}=3$) and that it is recommended~\cite{Sigworth2015} to set $N_{\theta_1}N_{\theta_2}$ in the order of $n^2$ to maintain a precise estimation of projection angles, the proposed method offers an interesting improvement in runtime over projection matching.
}

\section{Experiments}
\label{sec:exp_conditions}

\subsection{Datasets}

We test our algorithm on two synthetic datasets. The first dataset corresponds to the Holliday junction complex (HJC) density map, while the second corresponds to the Human patched 1 (PTCH1) protein. For each dataset, we generate the synthetic ground truth $V$ from the submitted density map~\cite{Laxmikanthan2016,Qi2018}, along with the associated atomic model in the Protein Data Bank using Chimera~\cite{Chimera2004}. The sizes of the HJC and PTCH1 volumes used in our simulations are $(90\times 90 \times 90)$ and $(84 \times 84 \times 84)$, \textcolor{black}{with voxel sizes of $2.867 \AA$ and $1.8 \AA$}, respectively. \textcolor{black}{We also synthesize a higher resolution version of  HJC with size $124\times124\times124$ and voxel size $2 \AA$. The first two volumes are used in our proof of concept simulations; the last volume is used in an experiment that mimics more realistic cryo-EM conditions (section V.E).} 

From those ground truths, we then generate $P$ projection images according to the image-formation model provided in~\eqref{eq:image_form}. We sample the orientation space using $P$ points in an equi-distributed fashion over $\{(\theta_{1,p},\theta_{2,p})\}_{p=1}^P$. The in-plane rotations are generated by uniformly sampling $P$ points on a $[0, 2\pi)$ interval. To perform in-plane translations, we move the center of the projection images randomly by at most $m_\mathbf{t}$ pixels in either horizontal or vertical directions. In our experiments, we use at most \textcolor{black}{$20000$}  projection images to demonstrate the feasibility of our method. Finally, the projection images are corrupted with additive Gaussian noise with zero mean and variance $\sigma^2$. The average signal-to-noise ratio (SNR) across all projection images is then given by $\mathrm{SNR}_{\mathrm{data}} = \frac{1}{P} \sum_{p=1}^P \frac{\Vert \mathbf{g}_p^\star\Vert_2^2}{\sigma^2}$ where $\mathbf{g}_p^\star$ correspond to noiseless measurements.

\subsection{Initial Density Map, Orientations, and In-Plane Translations }

To generate an approximate density map from which to start the refinement procedure, we use the initial density map generated by 3D \textit{ab-initio} model in Relion~\cite{Relion2017}. For the projection orientations, we consider two possible initializations.
\begin{itemize}
    \item Model \textit{Init-1}: We add a zero-mean  random variable uniformly distributed in $[-e_{\theta},e_{\theta}]$ to the ground-truth orientations, \textit{i.e.}, ${\bm{\uptheta}}^{\mathrm{init}}_p = \bm{\uptheta}_p^{\mathrm{true}} + \bm{\upvarepsilon}_{\theta,p}$ where $\bm{\upvarepsilon}_{\theta,p,j} \sim \mathrm{Unif}(-e_{\theta},e_{\theta})$, $p \in \{1,\ldots,P\}$ and $j \in \{1,2,3\}$. 
    \item Model \textit{Init-2}: We use projection-matching (or another angular assignment method) to assign the initial projection orientations $\bm{\uptheta}^{\mathrm{init}}_p$, $p \in \{1,\ldots,P \}$. \textcolor{black}{For this initialization, we use the angular assignments from the 3D ab-initio modeling in Relion.}
\end{itemize}

The in-plane translations $\{\mathbf{t}_p\}_{p=1}^P$ are all initialized by zeros. With our notations we have that $\bm{\Theta}^\mathrm{init} = \{\bm{\uptheta}_p^\mathrm{init} \}_{p=1}^P$, $\bm{\Theta}^\mathrm{true} = \{\bm{\uptheta}_p^\mathrm{true} \}_{p=1}^P$, $\bm{\Gamma}^\mathrm{init} = \{\mathbf{0} \}_{p=1}^P$, and  $\bm{\Gamma}^\mathrm{true} = \{\mathbf{t}_p^\mathrm{true} \}_{p=1}^P$.

\subsection{Tuning of the Hyper Parameters }
The parameters that need tuning are  $\lambda$, $\rho$, and $K_\mathrm{ADMM}$, as used in the update of the density map (Algorithm \ref{alg:volume_update}), and $\alpha_{\bm{\theta}}$, $\alpha_{\bm{t}}$, and $K_\mathrm{GD}$, as introduced in the update of the latent variables (Algorithm \ref{alg:angular_assignment}). In our experiments, we use $K_\mathrm{ADMM} = 2$ or $K_\mathrm{ADMM} = 5$, along with $K_\mathrm{GD} = 3$, $\alpha_{\bm{\uptheta}} = 10^{-7}$, $\alpha_{\bm{t}} = 10^{-5}$, and $\eta = 0.25$. The parameters $\lambda$ and $\rho$ grow like $\sigma$. We use the same set of parameters for the two molecules. Similar to \cite{donati2018fast}, the parameters of the KBWF used in the expansion of the volume in \eqref{eq:sampling}-\eqref{eq:KBWF} are $a=4$, $\alpha=19$, and $m=2$.

\subsection{Metrics}

To assess the quality of reconstruction, we use the Fourier shell correlation (FSC) between the reconstructed volume $
V^\mathrm{rec}$ and the ground-truth $V^\mathrm{gt}$, as defined by
\begin{align}
\mathrm{FSC}(r) = \frac{\sum\limits_{r_i \in \textbf{r}} \hat{V}^\mathrm{rec}(r_i) \hat{V}^\mathrm{gt}(r_i)^*}{\sqrt{\sum\limits_{r_i \in \textbf{r}} \vert \hat{V}^\mathrm{rec}(r_i) \vert^2 \sum\limits_{r_i \in \textbf{r}} \vert \hat{V}^\mathrm{gt}(r_i) \vert^2} }.
\end{align}
where $\textbf{r} = \{(x_i, y_i, z_i): |\sqrt{x_i^2+y_i^2+z_i^2} - r |\leq \varepsilon_r\}$, for $\varepsilon_r >0$, denotes the set of all points in the discrete Fourier domain that lie in a  spherical shell with inner radius $r-\varepsilon_r$ and outer radius $r+\varepsilon_r$, centered at origin. The FSC thus computes the correlation between two corresponding spherical shells of the density maps in the Fourier domain. Moreover, we use the SNR metric defined as $\mathrm{SNR}(V^\mathrm{gt},V^\mathrm{rec})=20 \log_{10} \frac{\Vert V^\mathrm{gt} \Vert_2}{\Vert V^\mathrm{gt}-V^\mathrm{rec} \Vert_2}$.

To assess the quality of the 3D orientation refinements, we visualize the deviations of the refined angles from their ground-truth values. In other words, we examine the distribution of $\{\theta_{i,p}^{\mathrm{true}}-\theta_{i,p}^{\mathrm{rec}}\}_{p=1}^P$ for $i \in \{1,2,3\}$ and compare it to $\{\theta_{i,p}^{\mathrm{true}}-\theta_{i,p}^{\mathrm{init}}\}_{p=1}^P$. When the difference between the angles is small (up to some global rotations), the distribution of the differences is more concentrated around zero. On the contrary, the distribution is more expanded for angles that are further away from their ground-truth values.

\begin{figure*}[t]
\begin{minipage}{1\textwidth}
        \includegraphics[width=1 \linewidth]{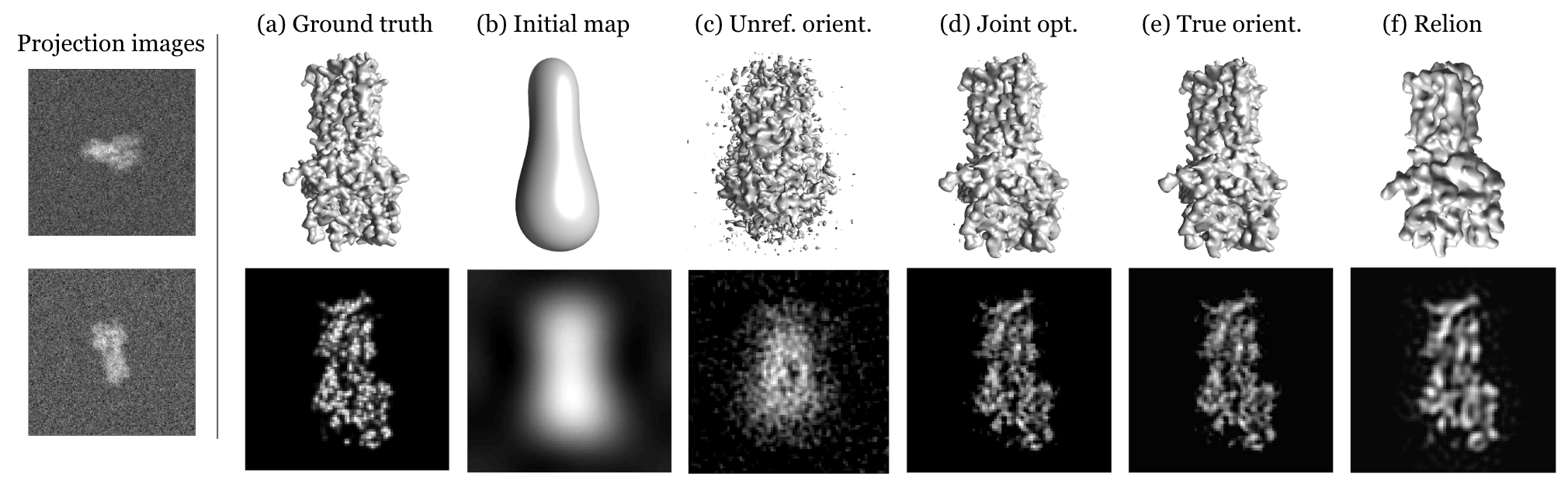}
        \subcaption[first caption.]{PTCH1, $P=500$, $f_c = 0.03$, $\mathrm{SNR}_\mathrm{data} = 3.5781 \,(\textrm{dB})$, $m_{\bf{t}}=0$.}
\end{minipage}
\vfill
\vspace{0.5cm}
    \begin{minipage}{1\textwidth}
    \includegraphics[width=\hsize]{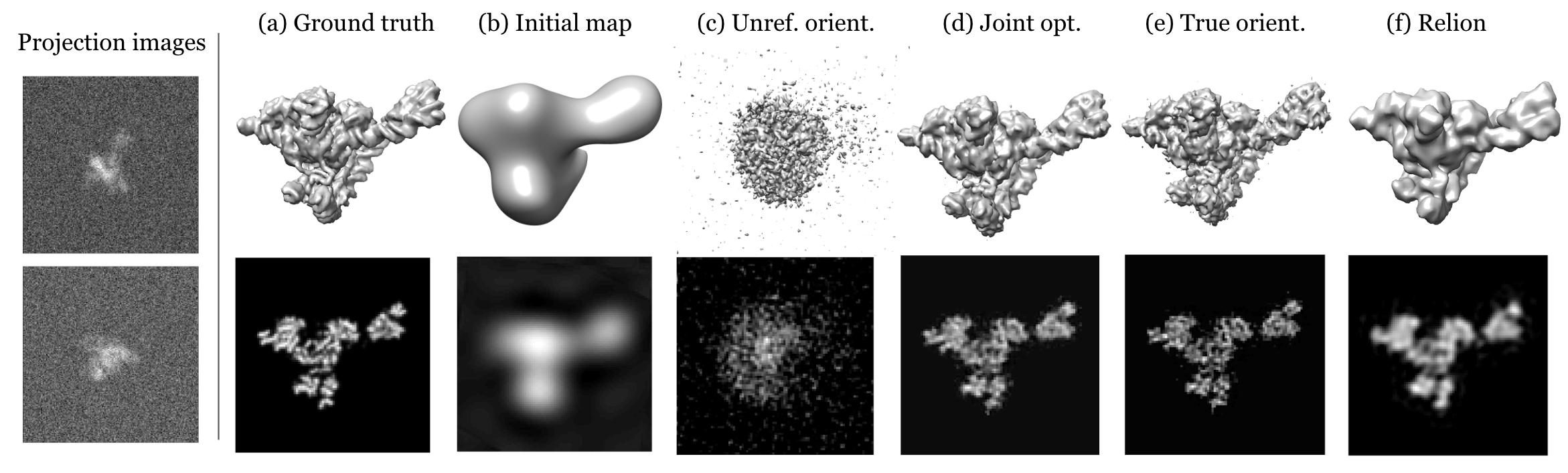}
    \subcaption[first caption.]{HJC, $P=500$, $f_c = 0.02$,  $\mathrm{SNR}_\mathrm{data} = -0.5733 \,(\textrm{dB})$, $m_{\bf{t}} = 3$.}
\end{minipage}
\vfill
    \caption{Reconstructions of PTCH1 and HJC. {Left}: Samples of the noisy projection images.  Top row: 3D structures. Bottom row: Intensity maps of the central slice of the structures. The presented volumes are (a) the ground truth, (b) the initial map, (c) the reconstruction with unrefined projection orientations, (d) the output of our joint refinement approach, (e) the reconstruction with the true projection orientations, (f) the output of Relion \textcolor{black}{after post-processing}. For both experiments, the latent variables are initialized following the \textit{Init-1} model with $e_{{\theta}} = 0.7$ [rad].}
    \label{fig:fullmodel}
\end{figure*}

\subsection{Compared Methods}

We compare our joint-optimization method to the following approaches:
\begin{enumerate}
\item \textit{Reconstruction with Unrefined Orientations}. We do not refine the initial angles and directly reconstruct the density map. This gives us an indication of the quality of reconstruction prior to the refinement procedure.  
\item \textit{Reconstruction with True Orientations}. We reconstruct the density map with the ground-truth orientations and in-plane translations. This serves as an oracle benchmark that allows us to quantify the improvement brought by our refinement procedure. 
\item \textit{Reconstruction with the Relion package}~\cite{Scheres2012}.
We run the \textit{3D auto-refine} function in Relion (version 2.1.0). The default parameters of this function (\textit{e.g.}, \textit{Initial angular sampling} and \textit{Local searches from auto sampling}) are used. For the particular experiments in which the in-plane translations are zero, the \textit{Initial offset range} and \textit{Initial offset step} parameters are set to their minimum values, which are $0$ and $0.1$, respectively. Otherwise, they are set to $4$ and $0.5$, respectively. Note that, to reduce the impact of noise when using Relion, we mask the projection images with a soft circular mask of a diameter that is proportional to the support of the density map.
\end{enumerate}

\textcolor{black}{All reconstructions from Relion are post-processed. We first apply a tight soft mask that embeds the maps. We then low-pass filter the volumes with a cut-off frequency that corresponds to the gold-standard FSC between the two half-maps; this is done using the post-processing function in Relion.}


\label{sec:results}

\section{Results}

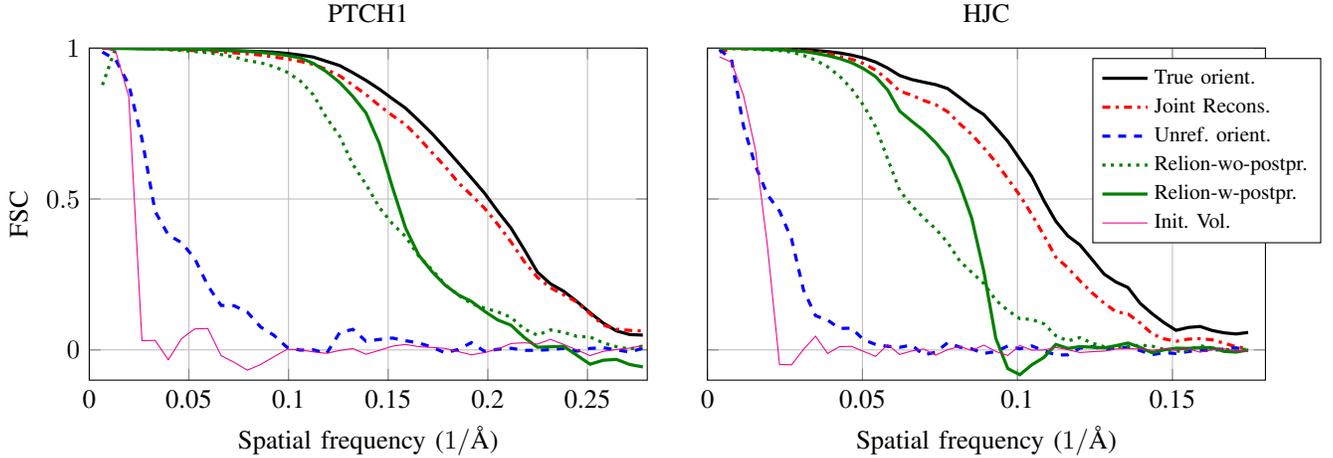
\begin{figure*}
\centering
	\begin{tikzpicture}
		\begin{groupplot}[group style={group size= 2 by 1, horizontal sep=0.8cm, vertical sep=2cm},     
	            	 legend pos= north east,        
					 legend style={at={(1.1,0.97)},legend cell align=left},
					 grid=both,                         
    				 height=6cm,width=9cm,
					 ymin=-0.1,ymax=1,xmin=0] 
		\nextgroupplot[xmin=0,xmax=0.28,xtick={0,0.05,0.1,0.15,0.2,0.25},xticklabels={0,0.05,0.1,0.15,0.2,0.25},title={PTCH1},ylabel={FSC},xlabel={Spatial frequency ($1/\AA$)},xtick distance=0.05] 	
			 \addplot[black,very thick] table[x=x, y=y, col sep=comma]{new_results/fig56_with_relion/fsc_true_7795.dat};	
			 \addplot[red,very thick,dashdotted] table[x=x, y=y, col sep=comma]{new_results/fig56_with_relion/fsc_ours_7795.dat};
			 \addplot[blue,very thick,dashed] table[x=x, y=y, col sep=comma]{new_results/fig56_with_relion/fsc_approx_7795.dat};
			 \addplot[darkgreen,very thick,dotted] table[x=x, y=y, col sep=comma]{new_results/fig56_with_relion/fsc_relion_nopost_7795.dat};
			 \addplot[darkgreen,very thick] table[x=x, y=y, col sep=comma]{new_results/fig56_with_relion/fsc_relion_wpost_7795.dat};
			 \addplot[magenta] table[x=x, y=y, col sep=comma]{new_results/fig56_with_relion/fsc_init_7795.dat};
		\nextgroupplot[xmin=0,xmax=0.18,xtick={0,0.05,0.1,0.15},xticklabels={0,0.05,0.1,0.15},yticklabels={,,},title={HJC},xlabel={Spatial frequency ($1/\AA$)},xtick distance=0.05] 	
			 \addplot[black,very thick] table[x=x, y=y, col sep=comma]{new_results/fig56_with_relion/fsc_true.dat};
			 \addplot[red,very thick,dashdotted] table[x=x, y=y, col sep=comma]{new_results/fig56_with_relion/fsc_ours.dat};
			 \addplot[blue,very thick,dashed] table[x=x, y=y, col sep=comma] {new_results/fig56_with_relion/fsc_approx.dat};
			 \addplot[darkgreen,very thick,dotted] table[x=x, y=y, col sep=comma] {new_results/fig56_with_relion/fsc_relion_nopost.dat};
			 \addplot[darkgreen,very thick] table[x=x, y=y, col sep=comma] {new_results/fig56_with_relion/fsc_relion_wpost.dat};
			 \addplot[magenta] table[x=x, y=y, col sep=comma] {new_results/fig56_with_relion/fsc_init.dat};	
			 \legend{{\footnotesize True orient.},{\footnotesize Joint Recons.},{\footnotesize Unref. orient.},{\footnotesize Relion-wo-postpr.},{\footnotesize Relion-w-postpr.},{\footnotesize Init. Vol.}};
		\end{groupplot}
	\end{tikzpicture}
	\caption{Comparison between the FSC of the density maps obtained from several baselines and the ground-truth density map. \textcolor{black}{Relion-w-postpr (greed solid curve) and Relion-wo-postpr (green dashed curve) refer to the Relion results with and without post-processing, respectively. Note that Relion-wo-postpr is obtained after averaging the two half-maps.} The experimental setups  are identical to the ones used in Figure \ref{fig:fullmodel}.}
    \label{fig:FSC_fullmodels}
\end{figure*}

\subsection{Visual Comparison}

We compare in Figure \ref{fig:fullmodel} the refined maps obtained using our join optimization scheme (Figure \ref{fig:fullmodel}-(d)) and the other methods (Figure \ref{fig:fullmodel}-(c,e,f)). In that experiment, the latent variables are initialized following the \textit{Init-1} model. 

As expected, the reconstruction fails when the unrefined 3D orientations are used (Figure~\ref{fig:fullmodel}-(c)). This confirms that angular refinement is required to achieve a successful reconstruction. Predictable as well is the fact that a perfect knowledge of the true 3D poses leads to a successful reconstruction (Figure~\ref{fig:fullmodel}-(e)).
It can be clearly seen that the results of our method (Figure \ref{fig:fullmodel}-(d)) closely resemble the reconstructed map resulting from perfect knowledge of the latent variables. This shows the ability of our method to appropriately refine the density map and the latent variables.

Figure~\ref{fig:fullmodel}-(f) contains the 3D density map refined by Relion. 
We observe that the map refined through our method is more similar to the ground-truth density map than the Relion output.

\subsection{FSC Curves}

The FSC curves of the reconstructed maps are sketched in Figure~\ref{fig:FSC_fullmodels}. These curves confirm that our joint-optimization approach (red dash-dotted curve) is able to appropriately refine the low-resolution initial map. Its performance indeed closely approaches that of the reconstruction with perfect knowledge of 3D orientations (solid curve). Moreover, our framework outperforms the Relion outcome with and without post-processing (green curves).


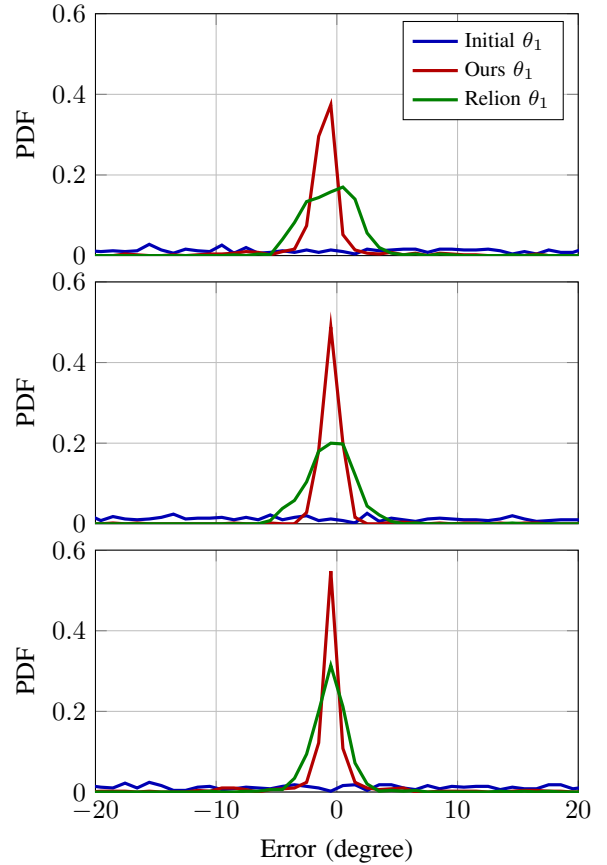
\begin{figure}
\centering
	\begin{tikzpicture}
		\begin{groupplot}[group style={group size= 1 by 3,                      
    				horizontal sep=0.8cm, vertical sep=0.35cm},     
	            	 legend pos= north east,        
					 legend style={legend cell align=left},
					 grid=both,                         
    				 height=4.8cm,width=8cm,
    				 xmin=-20,xmax=20,
					 ymin=0,ymax=0.6] 
		\nextgroupplot[ylabel={PDF},xticklabels={,,}] 	
			 \addplot[darkblue,very thick] table[x=x, y=y, col sep=comma]{new_results/fig56_with_relion/init_rot_3400.dat};	
			 \addplot[darkred,very thick] table[x=x, y=y, col sep=comma]{new_results/fig56_with_relion/ours_rot_3400.dat};
			 \addplot[darkgreen,very thick] table[x=x, y=y, col sep=comma]{new_results/fig56_with_relion/relion_rot_3400.dat};
			 \legend{{\footnotesize Initial $\theta_1$ },{\footnotesize Ours $\theta_1$ },{\footnotesize Relion $\theta_1$ }};
		\nextgroupplot[xticklabels={,,},ylabel={PDF}] 	
			 \addplot[darkblue,very thick] table[x=x, y=y, col sep=comma]{new_results/fig56_with_relion/init_tilt_3400.dat};	
			 \addplot[darkred,very thick] table[x=x, y=y, col sep=comma]{new_results/fig56_with_relion/ours_tilt_3400.dat};
			 \addplot[darkgreen,very thick] table[x=x, y=y, col sep=comma]{new_results/fig56_with_relion/relion_tilt_3400.dat};
		\nextgroupplot[xlabel={Error (degree)},ylabel={PDF}] 	
			 \addplot[darkblue,very thick] table[x=x, y=y, col sep=comma]{new_results/fig56_with_relion/init_psi_3400.dat};	
			 \addplot[darkred,very thick] table[x=x, y=y, col sep=comma]{new_results/fig56_with_relion/ours_psi_3400.dat};
			 \addplot[darkgreen,very thick] table[x=x, y=y, col sep=comma]{new_results/fig56_with_relion/relion_psi_3400.dat};
		\end{groupplot}
	\end{tikzpicture}
\caption{\color{black}Probability density function (PDF) of the differences between the true and refined projection orientations by our method $\{\theta_{i,p}^{\mathrm{true}}-\theta_{i,p}^{\mathrm{rec}}\}_{p=1}^P$ (red  curves), the true and refined projection orientations by Relion $\{\theta_{i,p}^{\mathrm{true}}-\theta_{i,p}^{\mathrm{Relion}}\}_{p=1}^P$ (green curves), as well as the true and the initial projection orientations $\{\theta_{i,p}^{\mathrm{true}}-\theta_{i,p}^{\mathrm{init}}\}_{p=1}^P$ (blue curves). The experimental setup is identical to Figure \ref{fig:fullmodel} (HJC). The x-axis is truncated between -20 and 20 degrees for the sake of clarity.}
\label{fig:angle_refined_hjc}
\end{figure}

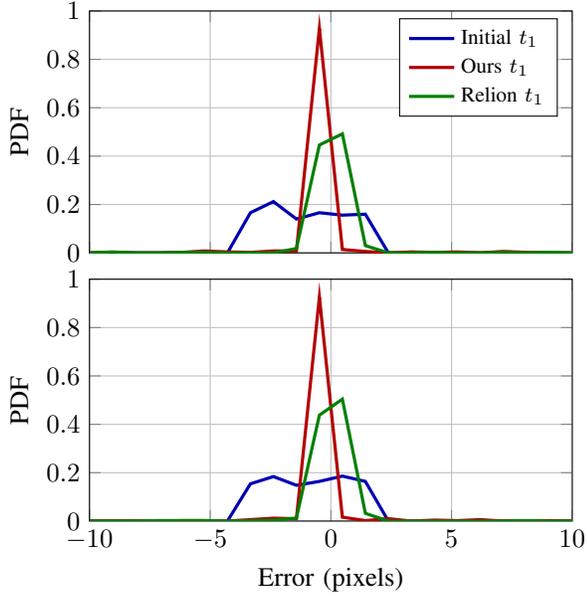
\begin{figure}
    \centering
	\begin{tikzpicture}
		\begin{groupplot}[group style={group size= 1 by 2,                      
    				horizontal sep=0.5cm, vertical sep=0.35cm},     
	            	 legend pos= north east,        
					 legend style={legend cell align=left},
					 grid=both,                         
    				 height=4.8cm,width=8cm,
    				 xmin=-10,xmax=10,
					 ymin=0,ymax=1] 
		\nextgroupplot[ylabel={PDF},xticklabels={,,}] 	
			 \addplot[darkblue,very thick] table[x=x, y=y, col sep=comma]{new_results/fig56_with_relion/init_shiftx_3400.dat};	
			 \addplot[darkred,very thick] table[x=x, y=y, col sep=comma]{new_results/fig56_with_relion/ours_shiftx_3400.dat};
			 \addplot[darkgreen,very thick] table[x=x, y=y, col sep=comma]{new_results/fig56_with_relion/relion_shiftx_3400.dat};
			 \legend{{\footnotesize Initial $t_1$ },{\footnotesize Ours $t_1$ },{\footnotesize Relion $t_1$ }};
		\nextgroupplot[ylabel={PDF},xlabel={Error (pixels)}] 	
			 \addplot[darkblue,very thick] table[x=x, y=y, col sep=comma]{new_results/fig56_with_relion/init_shifty_3400.dat};	
			 \addplot[darkred,very thick] table[x=x, y=y, col sep=comma]{new_results/fig56_with_relion/ours_shifty_3400.dat};
			 \addplot[darkgreen,very thick] table[x=x, y=y, col sep=comma]{new_results/fig56_with_relion/relion_shifty_3400.dat};
		\end{groupplot}
	\end{tikzpicture}
    \caption{\color{black} Probability density function (PDF) of the differences between the true and refined in-plane translations by our method $\{\mathbf{t}_{1,p}^\mathrm{true} - \mathbf{t}_{1,p}^\mathrm{rec}\}_{p=1}^P$ (red  curves), the true and refined in-plane translations by Relion $\{\mathbf{t}_{1,p}^\mathrm{true} - \mathbf{t}_{1,p}^\mathrm{Relion}\}_{p=1}^P$ (green curves), as well as the true and the initial in-plane translations $\{\mathbf{t}_{1,p}^\mathrm{true} - \mathbf{t}_{1,p}^\mathrm{init}\}_{p=1}^P$ (blue curves). The experimental setup is identical to Figure \ref{fig:fullmodel} (HJC).}
    \label{fig:shifts_refined_hjc}
\end{figure}

\subsection{Quality of Angular Refinement}
In Figure \ref{fig:angle_refined_hjc}, one finds the probability density function (PDF) of the differences between 1) the true and initial projection orientations (blue curve), 2) the true and refined projection orientations by our method (red curve), and 3) the true and refined projection orientations by Relion (green curve). The optimal PDF is obtained when all the differences are zero, up to a global rotation. The corresponding curve resembles a \textcolor{black}{delta} function that is one at zero, and zero elsewhere. Based on this, we observe that our proposed method performs well in recovering the projection orientations \textcolor{black}{and outperforms Relion}.

\textcolor{black}{Figure \ref{fig:shifts_refined_hjc} compares the PDF of the difference between 1) the true and initial in-plane translations (blue curve), 2) the true and refined in-plane translations by our method (red curve), and 3) the true and refined in-plane translations by Relion (green curve). Here as well, the figure demonstrates the ability of our method to refine in-plane translations, and its increased performance compared to Relion.}


\subsection{Convergence Results}

\begin{figure}
    \centering
    \includegraphics[width = 1 \linewidth]{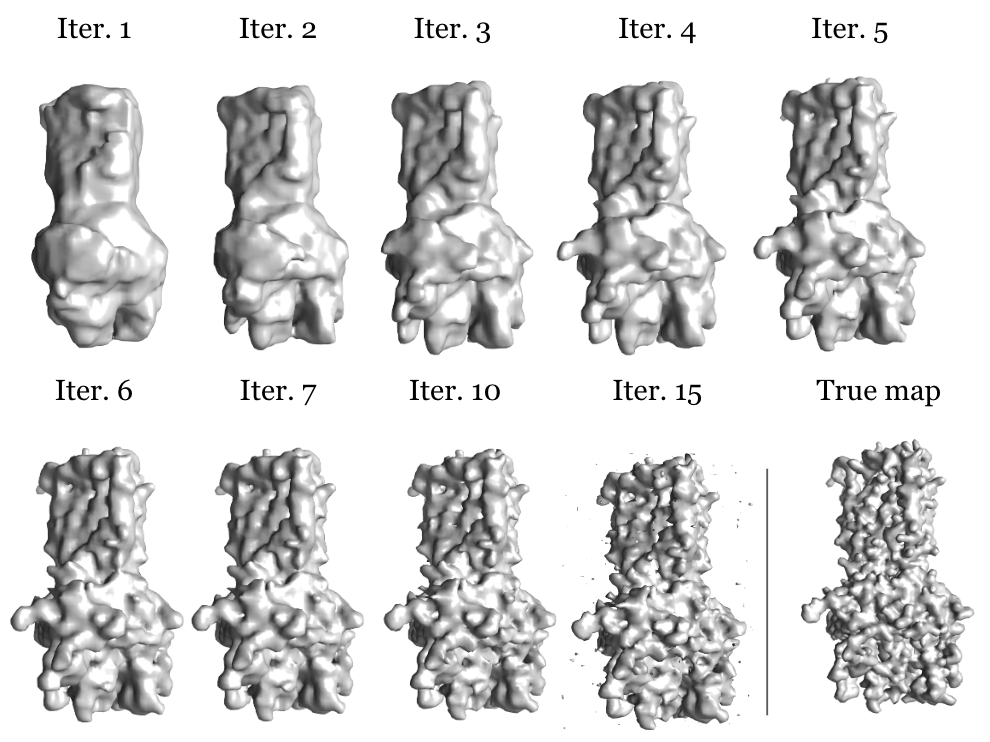}
    \caption{Evolution through iterations for the refinement of PTCH1. The experimental setup is the same as in  Figure \ref{fig:fullmodel}.}
    \label{fig:alg_evolution}
\end{figure}

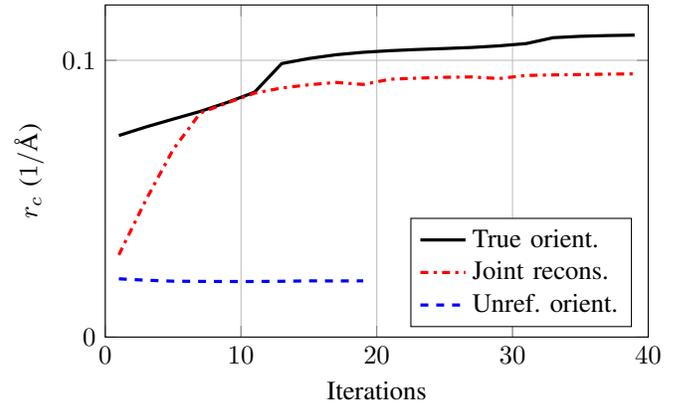
\begin{figure}
    \centering
    		\begin{tikzpicture}
		\begin{axis}[legend pos= south east,        
					 legend style={legend cell align=left},
					 grid=both,                         
				     set layers,cell picture=true,
					 ytick={0,0.1},
    				 height=6cm,width=8.8cm,
    				 xmin=0,xmax=40,
					 ymin=0,ymax=0.12,ylabel={$r_c$ ($1/\AA$)},xlabel=Iterations] 
					 \addplot[black,very thick] table[x=x, y=y, col sep=comma]{figures/3400_unif/evol_true_pixel.dat};	
					 \addplot[red,very thick,dashdotted] table[x=x, y=y, col sep=comma]{figures/3400_unif/evol_joint_pixel.dat};
					 \addplot[blue,very thick,dashed] table[x=x, y=y, col sep=comma]{figures/3400_unif/evol_approx_pixel.dat};	
			\legend{{True orient.},{Joint recons.},{Unref. orient.}};
		\end{axis}
	\end{tikzpicture}
    \caption{Evolution of $r_c$, the radial frequency at which the FSC equals $0.5$. Dash-dotted curve:  Our joint-reconstruction framework.  Solid curve: Reconstruction with true orientation projections. Dashed  curve: Reconstruction with unrefined orientation projections. The experimental setup is identical to Figure \ref{fig:fullmodel} (HJC). }
    \label{fig:fsc_evol}
\end{figure}

The evolution of the density map during refinement  is presented in Figure \ref{fig:alg_evolution}. The convergence in terms of resolution of our framework and  of two other baselines are shown in Figure~\ref{fig:fsc_evol}, where $r_c$ marks the radial frequency at which the FSC between the true and the reconstructed density map equals~$0.5$.

When the 3D projection orientations are perfectly known (solid curve), the reconstruction process achieves a high-resolution map in {\color{black} twenty} iterations. A key result is that our framework (dash-dotted curve) is able to converge to an almost equally-high resolution map starting from less-than-ideal 3D projection orientations. Once again, we observe failure when reconstructing with the unrefined set of projection orientations. This further confirms that refinement of the latent variables is vital to achieve a high-quality reconstruction of the map.  

\begin{figure*}
    \centering
    \includegraphics[width=1 \linewidth]{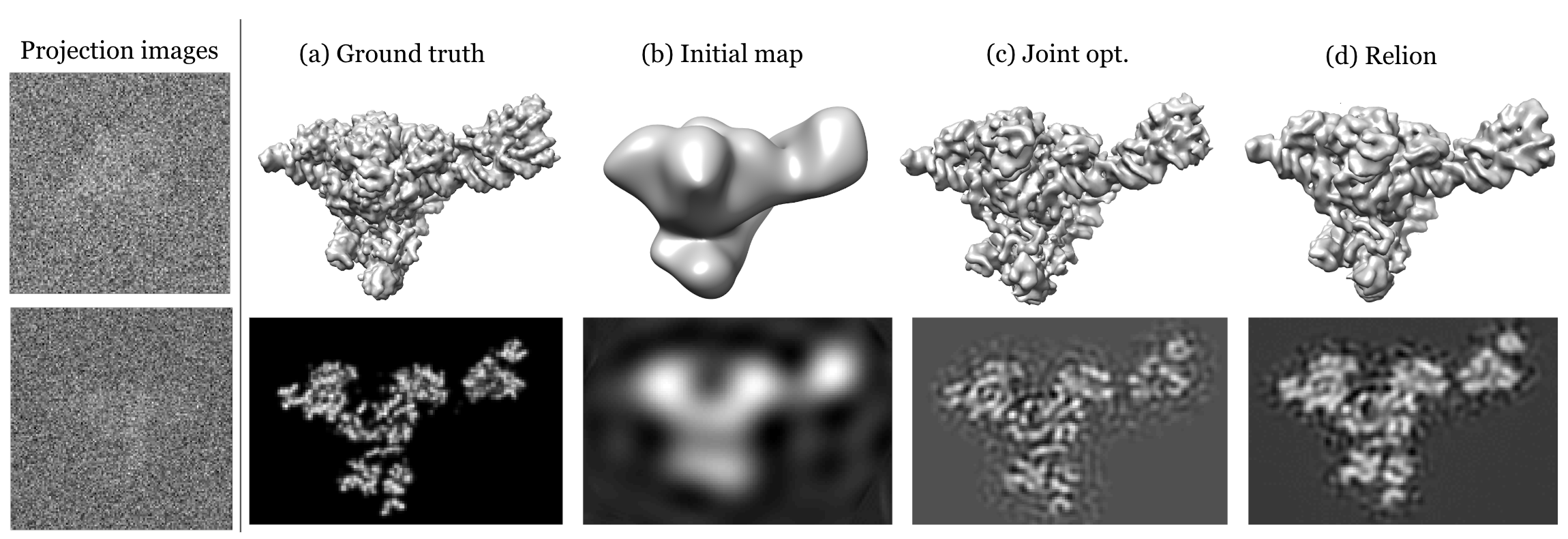}
    \caption{\color{black} Reconstructions of HJC. {Left}: Samples of the noisy projection images.  Top row: 3D structures. Bottom row: Intensity maps of the central slice of the structures. The presented volumes are (a) the ground truth, (b) the initial map, (c) the output of our joint refinement approach after post-processing, (d) the output of Relion after post-processing. For this experiments, the latent variables are initialized following the \textit{Init-2} model using the results from Relion 3D ab-initio modeling. The parameters of this experiment are: $P=20000$, $\mathrm{SNR}_{\mathrm{data}} = -14.2 \, \textrm{dB}$, $m_{\textbf{t}} = 3$.
    }
    \label{fig:3400_viz_1e4}
\end{figure*}


    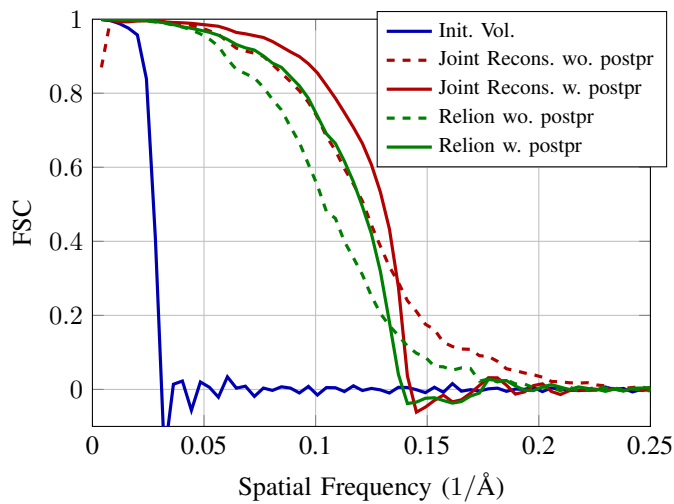
\begin{figure}
    \centering
	\begin{tikzpicture}
		\begin{groupplot}[group style={group size= 1 by 1,                      
    				horizontal sep=0.5cm, vertical sep=0.5cm},     
	            	 legend pos= north east,        
					 legend style={at={(1.03,1.02)},legend cell align=left},
					 grid=both,                         
    				 height=7cm,width=9cm,
    				 xticklabels={-1,0,0.05,0.1,0.15,0.2,0.25},
    				 xmin=0,xmax=0.25,
					 ymin=-.1,ymax=1] 
		\nextgroupplot[xlabel={Spatial Frequency ($1/\AA$)},ylabel={FSC}] 	
			 \addplot[darkblue,very thick] table[x=x, y=y, col sep=comma]{figures/3400_2e4/fsc_init.dat};	
			 \addplot[darkred,very thick,dashed] table[x=x, y=y, col sep=comma]{figures/3400_2e4/fsc_ours_nopost.dat};
			 \addplot[darkred,very thick] table[x=x, y=y, col sep=comma]{figures/3400_2e4/fsc_ours_wpost.dat};
			 \addplot[darkgreen,very thick,dashed] table[x=x, y=y, col sep=comma]{figures/3400_2e4/fsc_relion_nopost.dat};
			 \addplot[darkgreen,very thick] table[x=x, y=y, col sep=comma]{figures/3400_2e4/fsc_relion_wpost.dat};
			 \legend{{\footnotesize Init. Vol. },{\footnotesize Joint Recons. wo. postpr},{\footnotesize Joint Recons. w. postpr},{\footnotesize Relion wo. postpr },{\footnotesize Relion w. postpr }};
		\end{groupplot}
	\end{tikzpicture}
    \caption{Comparison between the FSC of the density maps obtained from several baselines and the ground-truth density map. \textcolor{black}{Relion-w-postpr (green solid curve) and Relion-wo-postpr (green dashed curve) refer to the Relion results with and without post-processing, respectively. Note that, Relion-wo-postpr is obtained after averaging the two half-maps.} The experimental setups  are identical to the ones used in Figure \ref{fig:3400_viz_1e4}.}
    \label{fig:fsc_2e4}
    \end{figure}

\subsection{Simulation of a Real Scenario}
\textcolor{black}{We then mimic a real scenario in which the output of the 3D \textit{ab initio} method provided by Relion  is used to initialize both the density map and the projection orientations. The in-plane translations are initialized with zeros. We use the same HJC structure to synthesize a volume with size $124\times124\times124$ and with a voxel size of $2 \AA$. The number of projection images is $20,000$ and the average SNR of the projection images is $-14.2 \textrm{dB}$. Examples of projection images are provided in Fig.~\ref{fig:3400_viz_1e4} (left most column).}

We  split the projection dataset in two halves and refine each half separately, starting from the same initial volume. Independent refinement of the two halves is a common approach in practice and has two main goals. First, by comparing the two refined half maps against one another, a convergence criterion is obtained. More precisely, we stop the refinement when the FSC between the two half maps fails to improve from one iteration to the next. Second, it reduces overfitting, especially in high-noise regimes. 

\textcolor{black}{A visual comparison of the density maps refined by our method and by Relion is presented in Fig.~\ref{fig:3400_viz_1e4} (c)-(e). Both results are post-processed by combining the half-maps and filtering out frequencies beyond the gold-standard FSC by applying a soft tight mask.}


\textcolor{black}{A quantitative comparison between our method and Relion based on FSC is provided in Fig.~\ref{fig:fsc_2e4}. Our method outperforms Relion both with and without post processing.}

\begin{figure}
    \centering
	\begin{tikzpicture}
		\begin{groupplot}[group style={group size= 1 by 3,                      
    				horizontal sep=0.5cm, vertical sep=0.35cm},     
	            	 legend pos= north east,        
					 legend style={legend cell align=left},
					 grid=both,                         
    				 height=4.8cm,width=8cm,
    				 xmin=-20,xmax=20,
					 ymin=0,ymax=0.25] 
		\nextgroupplot[ylabel={PDF},xticklabels={,,}] 	
			 \addplot[darkblue,very thick] table[x=x, y=y, col sep=comma]{figures/3400_2e4/init_rot_2e4.dat};	
			 \addplot[darkred,very thick] table[x=x, y=y, col sep=comma]{figures/3400_2e4/ours_rot_2e4.dat};
			 \addplot[darkgreen,very thick] table[x=x, y=y, col sep=comma]{figures/3400_2e4/relion_rot.dat};
			 \legend{{\footnotesize Initial $\theta_1$ },{\footnotesize Ours $\theta_1$ },{\footnotesize Relion $\theta_1$ }};
		\nextgroupplot[ylabel={PDF},xticklabels={,,}] 	
			 \addplot[darkblue,very thick] table[x=x, y=y, col sep=comma]{figures/3400_2e4/init_tilt_2e4.dat};	
			 \addplot[darkred,very thick] table[x=x, y=y, col sep=comma]{figures/3400_2e4/ours_tilt_2e4.dat};
			 \addplot[darkgreen,very thick] table[x=x, y=y, col sep=comma]{figures/3400_2e4/relion_tilt.dat};
		\nextgroupplot[ylabel={PDF},xlabel={Error (degree)}] 	
			 \addplot[darkblue,very thick] table[x=x, y=y, col sep=comma]{figures/3400_2e4/init_psi_2e4.dat};	
			 \addplot[darkred,very thick] table[x=x, y=y, col sep=comma]{figures/3400_2e4/ours_psi_2e4.dat};
			 \addplot[darkgreen,very thick] table[x=x, y=y, col sep=comma]{figures/3400_2e4/relion_psi.dat};
		\end{groupplot}
	\end{tikzpicture}
    \caption{\textcolor{black}{Probability density function (PDF) of the differences between the true and refined projection orientations by our method $\{\theta_{i,p}^{\mathrm{true}}-\theta_{i,p}^{\mathrm{rec}}\}_{p=1}^P$ (red  curves), the true and refined projection orientations by Relion (green curves), as well as the true and the initial projection orientations $\{\theta_{i,p}^{\mathrm{true}}-\theta_{i,p}^{\mathrm{init}}\}_{p=1}^P$ (blue curves). The experimental setup is identical to Figure \ref{fig:3400_viz_1e4}.}}
    \label{fig:angle_refined_2e4}
    \end{figure}
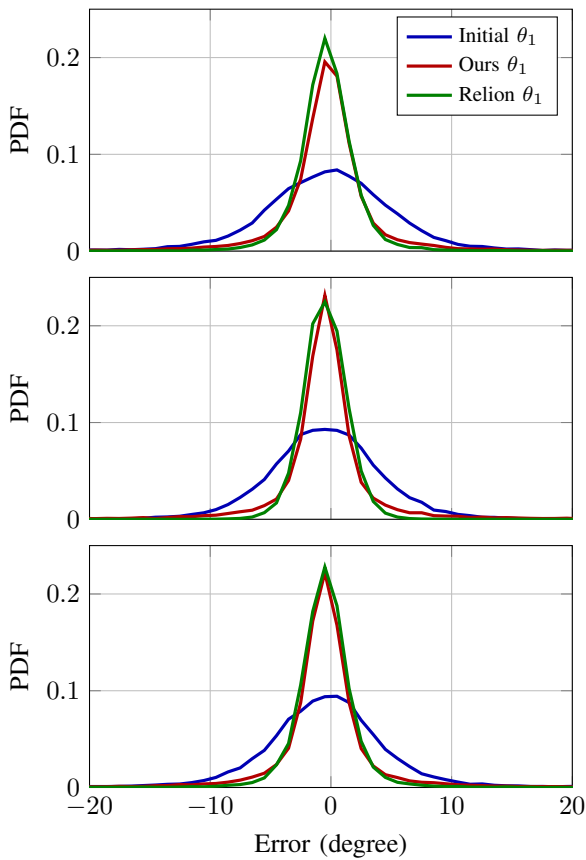

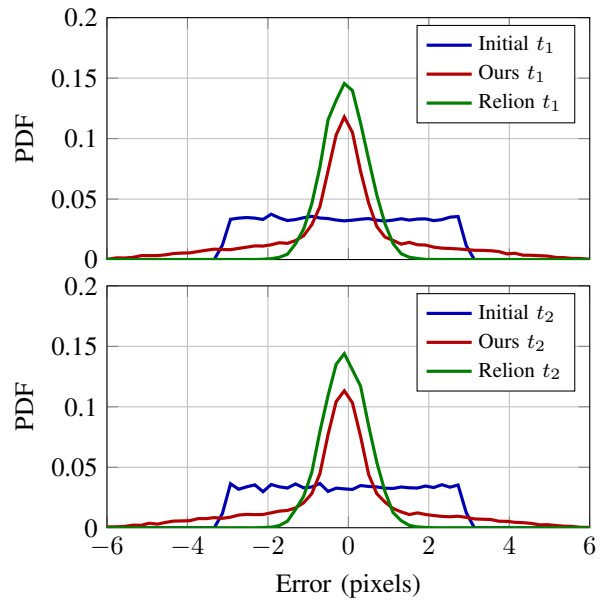
\begin{figure}
    \centering
	\begin{tikzpicture}
		\begin{groupplot}[group style={group size= 1 by 2,                      
    				horizontal sep=0.5cm, vertical sep=0.35cm},     
	            	 legend pos= north east,        
					 legend style={legend cell align=left},
					 grid=both,                         
    				 height=4.8cm,width=8cm,
    				 xmin=-6,xmax=6,
					 ymin=0,ymax=0.2] 
		\nextgroupplot[ylabel={PDF},yticklabels={-1,0,0.05,0.1,0.15,0.2},,xticklabels={,,}] 	
			 \addplot[darkblue,very thick] table[x=x, y=y, col sep=comma]{figures/3400_2e4/init_shiftx_2e4.dat};	
			 \addplot[darkred,very thick] table[x=x, y=y, col sep=comma]{figures/3400_2e4/ours_shiftx_2e4.dat};
			 \addplot[darkgreen,very thick] table[x=x, y=y, col sep=comma]{figures/3400_2e4/relion_shiftx_2e4.dat};
			 \legend{{\footnotesize Initial $t_1$ },{\footnotesize Ours $t_1$ },{\footnotesize Relion $t_1$ }};
		\nextgroupplot[ylabel={PDF},xlabel={Error (pixels)},yticklabels={-1,0,0.05,0.1,0.15,0.2}] 	
			 \addplot[darkblue,very thick] table[x=x, y=y, col sep=comma]{figures/3400_2e4/init_shifty_2e4.dat};	
			 \addplot[darkred,very thick] table[x=x, y=y, col sep=comma]{figures/3400_2e4/ours_shifty_2e4.dat};
			 \addplot[darkgreen,very thick] table[x=x, y=y, col sep=comma]{figures/3400_2e4/relion_shifty_2e4.dat};
			\legend{{\footnotesize Initial $t_2$ },{\footnotesize Ours $t_2$ },{\footnotesize Relion $t_2$ }};
		\end{groupplot}
	\end{tikzpicture}
    \caption{\textcolor{black}{Probability density function (PDF) of the differences between the true and refined in-plane translations by our method $\{\mathbf{t}_{1,p}^\mathrm{true} - \mathbf{t}_{1,p}^\mathrm{rec}\}_{p=1}^P$ (red  curves), the true and refined in-plane translations by Relion $\{\mathbf{t}_{1,p}^\mathrm{true} - \mathbf{t}_{1,p}^\mathrm{Relion}\}_{p=1}^P$ (green curves), as well as the true and the initial in-plane translations $\{\mathbf{t}_{1,p}^\mathrm{true} - \mathbf{t}_{1,p}^\mathrm{init}\}_{p=1}^P$ (blue curves). The experimental setup is identical to Figure \ref{fig:3400_viz_1e4}.}}
    \label{fig:shift_refined_2e4}
    \end{figure}

\textcolor{black}{To assess the quality of the refined latent variables, we compute the PDF of the errors between the refined and the ground-truth projection orientations and in-plane translations. We do this for both our method and Relion, and display the comparisons in Fig.~\ref{fig:angle_refined_2e4} and~\ref{fig:shift_refined_2e4}. We observe that the proposed method enjoys comparable performance with Relion for the refinement of the projection orientations and in-plane translations.}

\textcolor{black}{Hence, while the proposed angular refinement offers a substantial gain over Relion in the proof-of-concept experiments, the difference here is less significant for a larger volume and a noisier regime. However, our framework could be further improved in several ways. An option would be to add regularization for the latent variable estimation. The proposed framework could also be combined with the multi-scale approach proposed in~\cite{donati2018fast} to perform angle refinements at coarser scales, which demonstrates increased robustness to noise. We expect that those extensions would further improve the performance of the method while keeping an attractive numerical complexity as demonstrated in Section~\ref{sec:ComplexityComparaison}. These extensions are to be addressed in future works.}

\section{Conclusion}
\label{sec:conclusion}

We propose a variational 3D refinement framework for single-particle cryo-electron microscopy that jointly refines the density map and the 3D  projection orientations. The refinement of the orientations on the continuum does away with the computationally expensive projection-matching steps. We take alternating steps to update the density map and the latent variables. Steps of the classical method known as the alternating-direction method of multipliers are used to update the density map, while the latent variables are updated through gradient-descent. Our results demonstrate the ability of our framework to refine an approximate map from inaccurate 3D projection orientations. In addition, we show that the resolution of the refined map using our method closely approaches that of the map reconstructed with perfect knowledge of the 3D orientations.

\section{Acknowledgments}
This work was supported by the grant ERC-692726-GlobalBioIm and UIUC college of engineering strategic research initiatives. The authors would like to thank Dr. Masih Nilchian for its constructive feedback on the project. 

\appendix

\subsection{Fast Reconstruction with ADMM}
\label{subsec:appendix-ADMM}

We use the ADMM scheme proposed in~\cite{donati2018fast}. To that end, we introduce an auxiliary variable $\mathbf{u}$ and constrain its value by setting $\mathbf{u} = \mathbf{L} \mathbf{c}$, as done in~\eqref{eq:admm-minization-constrained}. The form of the augmented-Lagrangian function used in the ADMM procedure is thus given by 

\begin{multline}
\mathcal{L}_{\bm{\Theta},\bm{\Gamma}}(\textbf{c}, \mathbf{u}, \tilde{\mathbf{u}}) =  \frac{1}{2} \Vert \mathbf{g} - \mathbf{H}(\bm{\Theta}, \bm{\Gamma}) \, \mathbf{c}\Vert_2^2 + \lambda \mathcal{R}(\mathbf{u})  \\
 + \tilde{\mathbf{u}}^T (\mathbf{Lc} - \mathbf{u}) + \frac{\rho}{2} \Vert \mathbf{Lc} - \mathbf{u}\Vert_2^2, \label{eq:lag_def}
\end{multline}
where $\tilde{\mathbf{u}}$ is the Lagrangian multiplier that corresponds to the constraint $\mathbf{u} = \mathbf{Lc}$  and $\rho$ is the penalty parameter. Then, the ADMM algorithm alternates between a minimization of $\mathcal{L}$ with respect to
$\mathbf{u}$, a minimization of $\mathcal{L}$ with respect to $\mathbf{c}$, and an update of the dual variable $\tilde{\mathbf{u}}$.

The minimization of $\mathcal{L}$ with respect to $\mathbf{u}$ (Step \ref{algo:prox-u} in Algorithm~\ref{alg:volume_update}) results in  
\begin{align}
    \mathbf{u}^{k+1} = \textrm{arg} \min_{\mathbf{u}} \; \left( \lambda \mathcal{R}( \mathbf{u} ) + \frac{\rho}{2} \Vert \mathbf{L} \mathbf{c}^{k} - \mathbf{u} + {\tilde{\mathbf{u}}^k}/{\rho}\Vert_2^2 \right), \label{eq:z_update}
\end{align}
where one recognizes the proximity operator of $\mathcal{R}$. Hence, 
\begin{align}
    \mathbf{u}^{k+1} = \textrm{prox}_{\frac{\lambda}{\rho} \mathcal{R}} (\mathbf{Lc}^k - {\tilde{\mathbf{u}}^k}/{\rho}).
\end{align}

Then, the objective function involved in the update of $\mathbf{c}$ at  Line~\ref{algo:standard-admm-linear-step} of Algorithm \ref{alg:volume_update} is
\begin{multline}
      \mathcal{L}_{{\bm{\Theta}},{\bm{\Gamma}}}(\mathbf{c}, \mathbf{u}^{k+1}, \tilde{\mathbf{u}}^k) 
     =    \frac{1}{2} \Vert \mathbf{g} - \mathbf{H}(\bm{\Theta}, \mathbf{\Gamma}) \, \mathbf{c}\Vert_2^2 \\
     + \frac{\rho}{2} \| \mathbf{Lc} - \mathbf{u}^{k+1} + {\tilde{\mathbf{u}}^k}/{\rho} \|_2^2, \label{eq:c_update}
\end{multline}
which is a convex quadratic function of $\mathbf{c}$. Its minimization yields the linear system of equations
\begin{multline}
\left(  \mathbf{H}^T \mathbf{H}(\bm{\Theta}, \bm{\Gamma}) + \rho \mathbf{L}^T \mathbf{L} \right) \textbf{c}^{k+1} =  \rho \mathbf{L}^T ( \mathbf{u}^{k+1} -  \tilde{\mathbf{u}}^k/\rho) \\ + \mathbf{H}^T(\bm{\Theta}, \bm{\Gamma}) \mathbf{g}.
\end{multline}
We solve it in terms of $\mathbf{c}$ using a conjugate-gradient method. Note that, for the x-ray operator $\mathbf{H}(\bm{\Theta}, \bm{\Gamma})$, the quantity $ \mathbf{H}^T \mathbf{H}(\bm{\Theta}, \bm{\Gamma})\mathbf{c}$ can be efficiently computed at the cost of one FFT and one inverse FFT~\cite{delaney1996fast,vonesch2011fast,donati2018fast}.
{\color{black}
Indeed, we have that 
\begin{equation}
    \mathbf{H}^T\mathbf{H}(\bm{\Theta}, \bm{\Gamma})\mathbf{c} = \mathbf{w}({\bm{\Theta}}) \ast \mathbf{c},
\end{equation}
where the kernel $\mathbf{w}({\bm{\Theta}}) \in \mathbb{R}^N$ is given by, $\forall \mathbf{k} \in \Omega_\mathrm{3D},$
\begin{equation}\label{eq:kernelHtH}
     \left[\mathbf{w}({\bm{\Theta}})\right]_{\mathbf{k}} = \frac{1}{\mathrm{det}(\bm{\Lambda})}\sum_{p=1}^P \left( \psi_{\bm{\uptheta}_p} \ast \psi_{\bm{\uptheta}_p}^\vee \right) (\mathbf{M}_{\bm{\uptheta_p}^{\perp}} \mathbf{k}),
\end{equation}
with $\psi_{\bm{\uptheta}_p} = h \ast \mathcal{P}_{\bm{\uptheta}_p}(\varphi)$ a function that maps $\mathbb{R}^2$ to $\mathbb{R}$. It is worth to mention that the kernel $\mathbf{w}({\bm{\Theta}})$ does not depend on the in-plane translations $\bm{\Gamma}$. 

A similar strategy can be deployed to efficiently evaluate the quantity $\mathbf{H}^T(\bm{\Theta}, \bm{\Gamma}) \mathbf{g}$. Let $g_p$ be the continuous version of the measurements $\mathbf{g}_p$ (\ie, $g_p[\mathbf{m}] = g_p(\bm{\Lambda}\mathbf{m})$), which can for instance be obtained via some interpolation of the elements of $\mathbf{g}_p$. We then have that~\cite{donati2018fast}
\begin{equation}\label{eq:Htb}
    \left[\mathbf{H}^T(\bm{\Theta}, \bm{\Gamma}) \mathbf{g}\right]_\mathbf{k} \mkern-10mu = \frac{1}{\mathrm{det}(\bm{\Lambda})}\sum_{p=1}^P \left(g_p \ast \psi_{\bm{\uptheta}_p}^\vee \right)(\mathbf{M}_{\bm{\uptheta_p}^{\perp}} \mathbf{k} + \mathbf{t}_p).  
\end{equation}
The interest of~\eqref{eq:Htb} is that $g_p \ast \psi_{\bm{\uptheta}_p}^\vee$ can be precomputed on a fine grid using discrete 2D convolutions. Then, each term in the sum~\eqref{eq:Htb} comes at the price of an interpolation of this precomputed quantity.
}

Finally, the Lagrange multiplier $\tilde{\mathbf{u}}$ in ADMM is updated through a simple gradient-ascent step  (Step \ref{admmcg-update-u} in Algorithm~\ref{alg:volume_update}).

{
\color{black}
\subsection{Proof of Theorem~\ref{th:grad}} \label{sec:ProofGrad}

  Let us expand $\mathcal{J}_p(\bm{\uptheta},\mathbf{t})$ in~\eqref{eq:latentVarOpt-objFunction} as
  \begin{align}
     \mkern-12mu \mathcal{J}_p(\bm{\uptheta},\mathbf{t}) & = \frac12 \mathbf{c}^T \mathbf{H}^T \mathbf{H}(\bm{\uptheta},\mathbf{t}) \mathbf{c} - \mathbf{c}^T \mathbf{H}^T(\bm{\uptheta},\mathbf{t}) \mathbf{g}_p + \frac12 \|\mathbf{g}_p \|^2 \notag \\
      & = \frac12 \mathbf{c}^T (\mathbf{w}({\bm{\uptheta}}) \ast \mathbf{c}) - \mathbf{c}^T \mathbf{H}^T(\bm{\uptheta},\mathbf{t}) \mathbf{g}_p + \frac12 \|\mathbf{g}_p \|^2, \label{eq:proofTh1}
  \end{align}
where $\mathbf{w}({\bm{\uptheta}})$ corresponds to one term in the sum~\eqref{eq:kernelHtH}. Moreover,  because $\varphi$ is isotropic, we have that $\mathcal{P}_{\bm{\uptheta}}(\varphi)=\mathcal{P}(\varphi)$, a quantity that does not depend on $\bm{\uptheta}$. Hence, 
\begin{equation}
    \left[ \mathbf{w}({\bm{\uptheta}})\right]_{\mathbf{k}} =   \frac{1}{\mathrm{det}(\bm{\Lambda})}\left( \psi \ast \psi^\vee \right) (\mathbf{M}_{\bm{\uptheta}^{\perp}} \mathbf{k}), \label{eq:proofTh2}
\end{equation}
with $\psi = h * \mathcal{P}(\varphi)$.

Then, from~\eqref{eq:proofTh1}, one easily sees that, for all $v \in \{\theta_1,\theta_2,\theta_3,t_1,t_2\}$,
\begin{equation}
    \frac{\partial \mathcal{J}_p}{\partial v}(\bm{\uptheta},\mathbf{t}) = \frac12 \mathbf{c}^{T}\left(\mathbf{r}_v \ast \mathbf{c} - 2 \mathbf{q}_v \right),
\end{equation}
where
\begin{equation}
    \mathbf{r}_v = \frac{\partial \mathbf{w}(\bm{\uptheta})}{\partial v} \; \text{ and } \; \mathbf{q}_v = \frac{\partial  \mathbf{H}^T(\bm{\uptheta},\mathbf{t}) \mathbf{g}_p}{\partial v}.
\end{equation}
We now distinguish two cases.
\subsubsection{Case $v = \theta_i$ for $i \in \{1,2,3\}$}
From~\eqref{eq:proofTh2} and the chain rule, we get that
\begin{equation} \label{eq:proofTh4}
     r_v[\mathbf{k}] =   \frac{1}{\mathrm{det}(\bm{\Lambda})} \left( \frac{\partial \mathbf{M}_{{\bm{\uptheta}}^{\perp}} }{\partial \theta_i}   \mathbf{k} \right)^{T} \bm{\nabla} \left( \psi \ast \psi^\vee \right) (\mathbf{M}_{\bm{\uptheta}^{\perp}} \mathbf{k}),
\end{equation}
where $\frac{\partial  \mathbf{M}_{{\bm{\uptheta}}^{\perp}}}{\partial \theta_i} \in \mathbb{R}^{2 \times 3}$ contains the entry-wise derivatives with respect to $\theta_i$ of the matrix $\textbf{M}_{{\mathbf{\theta}}^{\perp}}$ given in~\eqref{eq:M_omega}.  Moreover, from the definition of $\psi : \mathbf{y} \mapsto ( h * \mathcal{P}(\varphi))(\mathbf{y})$, with $\mathbf{y}= (y_1,y_2) \in \mathbb{R}^2$, and from the derivation property of the convolution, we have that
\begin{align}
    \bm{\nabla}  \left( \psi \ast \psi^\vee \right)  & =
    \begin{pmatrix}
    \displaystyle \frac{\partial h }{\partial y_1} * \mathcal{P}(\varphi) \ast \psi^\vee \\
    \displaystyle  \frac{\partial h }{\partial y_2} * \mathcal{P}(\varphi)\ast \psi^\vee
    \end{pmatrix}  
     =
     \begin{pmatrix}
    \displaystyle  h * \frac{\partial \mathcal{P}(\varphi) }{\partial y_1} \ast \psi^\vee \\
    \displaystyle  h * \frac{\partial \mathcal{P}(\varphi) }{\partial y_2} \ast \psi^\vee
    \end{pmatrix}. \label{eq:GradPsiConvPsiVee}
\end{align}
Note that we could have also differentiated $\psi^\vee$ (instead of $h$ or $\mathcal{P}(\phi)$).

For $\mathbf{q}_v$, we get from~\eqref{eq:Htb} that
\begin{align}
    \mkern-8mu q_v[\mathbf{k}] &= \frac{1}{\mathrm{det}(\bm{\Lambda})}   \frac{\partial (g_p \ast \psi^\vee)(\mathbf{M}_{\bm{\uptheta}^{\perp}} \mathbf{k} + \mathbf{t})}{\partial \theta_i}  \notag \\
    & =   \frac{1}{\mathrm{det}(\bm{\Lambda})}  \bigg( \frac{\partial \mathbf{M}_{{\bm{\uptheta}}^{\perp}} }{\partial \theta_i}   \mathbf{k} \bigg)^{\mkern-3mu T}   \bm{\nabla}(g_p \ast \psi^\vee)  (\mathbf{M}_{\bm{\uptheta}^{\perp}} \mathbf{k}+\mathbf{t}), \label{eq:proofTh6}
\end{align}
where $\bm{\nabla}  \left( g_p \ast \psi^\vee \right) $ is obtained in the same way as~\eqref{eq:GradPsiConvPsiVee}, with differentiation on $\psi^\vee$ instead of $g_p$.

\subsubsection{Case $v =t_j$ for $j \in \{1,2\}$} As $\mathbf{w}(\bm{\uptheta})$ does not depend on the in-plane translation $\mathbf{t}$, we have that $\mathbf{r}_v = \mathbf{0}_{\mathbb{R}^N}$. For $\mathbf{q}_v$, as in~\eqref{eq:proofTh6}, we get that
\begin{align}
    q_v[\mathbf{k}] &= \frac{1}{\mathrm{det}(\bm{\Lambda})}   \frac{\partial (g_p \ast \psi^\vee)}{\partial y_j} (\mathbf{M}_{\bm{\uptheta}^{\perp}} \mathbf{k} + \mathbf{t}).
    \end{align}
}


\subsection{Proof of Proposition~\ref{propo:KBWF}} \label{sec:ProofKBWF}
The closed-form expression of the x-ray transform of the KBWF $\varphi$ in~\eqref{eq:KBWF} is provided in~\cite{Lewitt1990} as 
\begin{equation}\label{eq:proofPropo2-1}
     \mkern-8mu \mathcal{P}(\varphi)(\mathbf{y}) = \frac{a \sqrt{2\pi/\alpha}}{I_m(\alpha)} \, \mkern-5mu  \beta_a(\|\mathbf{y}\|)^{m+\frac12}  I_{m + \frac12} \mkern-5mu\left( \alpha \beta_a(\|\mathbf{y}\|) \right) ,
\end{equation}
where $\beta_a(r) = \sqrt{1-(r/a)^2}$ and $I_m$ is the modified Bessel function of order $m$. Now, let us introduce the function $f(u) = (\alpha u)^{m+\frac12} I_{m+\frac12}(\alpha u)$ whose derivative is $f'(u)= \alpha (\alpha u)^{m+\frac12} I_{m-\frac12}(\alpha u)$. Then, we can write~\eqref{eq:proofPropo2-1} as
\begin{equation}
    \mathcal{P}(\varphi)(\mathbf{y}) = \frac{a \sqrt{2\pi/\alpha}}{I_m(\alpha)} \ \frac{1}{\alpha^{m + \frac12}} f(\beta_a(\|\mathbf{y}\|))
\end{equation}
and, for all $v \in \{1,2\}$, obtain that
\begin{equation}\label{eq:proofPropo2-2}
    \frac{\partial \mathcal{P}(\varphi)}{\partial y_v}(\mathbf{y}) = \frac{a \sqrt{2\pi/\alpha}}{I_m(\alpha) \alpha^{m + \frac12}} \frac{y_v}{\|\mathbf{y}\|} \beta_a'(\|\mathbf{y}\|) f'(\beta_a(\|\mathbf{y}\|)).
\end{equation}
Finally, the injection of $f'$ and  $\beta_a'(r) = \left(-\frac{r}{a^2} \left( 1 - (r/a)^2 \right)^{-\frac12} \right)= \left(- \frac{r}{a^2\beta_a(r)} \right)$ into~\eqref{eq:proofPropo2-2} leads to
\begin{align}
      \frac{\partial \mathcal{P}(\varphi_{\alpha,a})}{\partial y_v}(\mathbf{y})  &= - \frac{a \sqrt{2\pi/\alpha}}{I_m(\alpha)\alpha^{m + \frac12}} \frac{y_v}{\|\mathbf{y}\|} \frac{\|\mathbf{y}\|\alpha (\alpha \beta_a(\|\mathbf{y}\|))^{m+\frac12}}{a^2\beta_a(\|\mathbf{y}\|)} 
       \notag \\
       & \quad \times   I_{m-\frac12}(\alpha \beta_a(\|\mathbf{y}\|)), \notag \\
       & =  - \frac{\alpha y_v \sqrt{2\pi/\alpha}}{a I_m(\alpha)}\beta_a(\|\mathbf{y}\|)^{m-\frac12}  \notag \\
        & \quad \times  I_{m-\frac12}(\alpha \beta_a(\|\mathbf{y}\|)),
\end{align}
which completes the proof.

\bibliographystyle{IEEEbib}

\bibliography{references.bib}

\end{document}